\documentclass[sigplan,screen]{acmart}

\copyrightyear{2025}
\acmYear{2025}
\setcopyright{rightsretained}
\acmConference[ASPLOS '25]{Proceedings of the 30th ACM International Conference on Architectural Support for Programming Languages and Operating Systems, Volume 1}{March 30--April 3, 2025}{Rotterdam, Netherlands}
\acmBooktitle{Proceedings of the 30th ACM International Conference on Architectural Support for Programming Languages and Operating Systems, Volume 1 (ASPLOS '25), March 30--April 3, 2025, Rotterdam, Netherlands}
\acmDOI{10.1145/3669940.3707215}
\acmISBN{979-8-4007-0698-1/25/03}

\makeatletter
\gdef\@copyrightpermission{
  \begin{minipage}{0.3\columnwidth}
   \href{https://creativecommons.org/licenses/by/4.0/}{\includegraphics[width=0.90\textwidth]{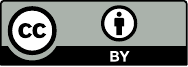}}
  \end{minipage}\hfill
  \begin{minipage}{0.7\columnwidth}
   \href{https://creativecommons.org/licenses/by/4.0/}{This work is licensed under a Creative Commons Attribution International 4.0 License.}
  \end{minipage}
  \vspace{5pt}
}
\makeatother

\usepackage{tikz}
\usepackage{amsmath}
\usepackage{xspace}
\usepackage{amsfonts}
\usepackage{subcaption}
\usepackage{booktabs}
\usepackage{multirow}
\usepackage{xcolor}
\usepackage{enumitem}
\usepackage{balance}

\newcommand{\sys}{Helix\xspace}

\newcommand{\camready}[1]{{#1}}

\renewcommand{\paragraph}[1]{\noindent\textbf{#1}}

\begin{document}

\setlist[itemize]{noitemsep}
\setlist[itemize]{leftmargin=*}

\title[\sys: Serving LLMs over Heterogeneous GPUs and Network via Max-Flow]{\sys: {Serving Large Language Models over Heterogeneous GPUs and Network via Max-Flow}}

\author{Yixuan Mei}
\email{yixuanm@andrew.cmu.edu}
\orcid{0009-0003-5781-9164}
\affiliation{
    \institution{Carnegie Mellon University}
    \city{Pittsburgh}
    \state{PA}
    \country{USA}
}

\author{Yonghao Zhuang}
\email{yzhuang2@andrew.cmu.edu}
\orcid{0009-0001-8969-7478}
\affiliation{
    \institution{Carnegie Mellon University}
    \city{Pittsburgh}
    \state{PA}
    \country{USA}
}

\author{Xupeng Miao}
\email{xupeng@cmu.edu}
\orcid{0000-0002-9371-8358}
\affiliation{
    \institution{Carnegie Mellon University}
    \city{Pittsburgh}
    \state{PA}
    \country{USA}
}

\author{Juncheng Yang}
\email{juncheny@cs.cmu.edu}
\orcid{0000-0002-0412-1139}
\affiliation{
    \institution{Carnegie Mellon University}
    \city{Pittsburgh}
    \state{PA}
    \country{USA}
}

\author{Zhihao Jia}
\email{zhihao@cmu.edu}
\orcid{0000-0002-1270-5185}
\affiliation{
    \institution{Carnegie Mellon University}
    \city{Pittsburgh}
    \state{PA}
    \country{USA}
}

\author{Rashmi Vinayak}
\email{rvinayak@cs.cmu.edu}
\orcid{0000-0002-2227-7460}
\affiliation{
    \institution{Carnegie Mellon University}
    \city{Pittsburgh}
    \state{PA}
    \country{USA}
}

\renewcommand{\shortauthors}{Yixuan Mei et al.}

\begin{abstract}
This paper introduces \sys, a distributed system for high-throughput, low-latency large language model (LLM) serving in heterogeneous GPU clusters.
The key idea behind \sys is to formulate inference computation of LLMs over heterogeneous GPUs and network connections as a {\em max-flow} problem on directed, weighted graphs, whose nodes represent GPU instances and edges capture both GPU and network heterogeneity through their capacities. 
\sys then uses a mixed integer linear programming (MILP) algorithm to discover highly optimized strategies to serve LLMs on heterogeneous GPUs.
This approach allows \sys to jointly optimize model placement and request scheduling, two highly entangled tasks in heterogeneous LLM serving.
Our evaluation on several heterogeneous clusters ranging from 24 to 42 GPU nodes shows that \sys improves serving throughput by up to 3.3$\times$ and reduces prompting and decoding latency by up to 66\% and 24\%, respectively, compared to existing approaches.
\sys is available at \url{https://github.com/Thesys-lab/Helix-ASPLOS25}.
\end{abstract}



\maketitle

\section{Introduction}
\label{sec1:intro}

\begin{table}
    \centering
    \caption{Minimum numbers of GPUs required to serve LLMs in existing homogeneous serving systems. We use half of GPU memory to store model parameters and the other half for key-value cache.
    \vspace{-0.8em}
    }
    \label{tab:gpu_requirement}
    \footnotesize
    \begin{tabular}{l|c|c|c|c}
    \toprule
    {\bf } & {\bf Num. of} & {\bf Num. of} & {\bf Num. of} & {\bf Num. of} \\
    {\bf LLMs} & {\bf Parameters} & {\bf L4s} & {\bf A100s} & {\bf H100s}\\
    \midrule
    LLaMA-2~\cite{touvron2023llama2} & 70 billion & 12 & 7 & 4\\
    GPT-3~\cite{achiam2023gpt} & 175 billion & 30 & 18 & 9 \\
    Grok-1~\cite{xaigrok} & 314 billion & 53 & 32 & 16 \\
    \camready{LLaMA-3~\cite{dubey2024llama}} & \camready{405 billion} & \camready{68} & \camready{41} & \camready{21}\\
    \bottomrule
    \end{tabular}
    \vspace{-0.4em}
\end{table}
\begin{table}
\centering
\caption{\camready{Availability of different GPU instances in 6 regions on Google Compute Engine~\cite{googlecloud2024compute}.}
\vspace{-0.8em}
}
\label{tab:sec1-cluster-heterogeneity}
\footnotesize
\begin{tabular}{@{}l|cccccc@{}}
\toprule
 & \multicolumn{6}{c}{\textbf{GPU Type}}              \\
          \textbf{Region}                & \textbf{H100} & \textbf{A100 80GB} & \textbf{A100 40GB} & \textbf{L4} & \textbf{T4} & \textbf{V100} \\ \midrule
us-central-1              & $\checkmark$   & $\checkmark$        & $\checkmark$        & $\checkmark$ & $\checkmark$ & $\checkmark$   \\
us-east-4                 & $\checkmark$   & $\checkmark$        & $\times$       & $\checkmark$ & $\checkmark$ & $\times$  \\
us-east-1                 & $\times$  & $\times$       & $\checkmark$        & $\checkmark$ & $\checkmark$ & $\checkmark$   \\
eu-west-3                 & $\checkmark$   & $\times$       & $\times$       & $\checkmark$ & $\checkmark$ & $\times$  \\
asia-ne-1                 & $\checkmark$   & $\times$       & $\checkmark$        & $\checkmark$ & $\checkmark$ & $\times$  \\
asia-ne-3                 & $\times$  & $\times$       & $\checkmark$        & $\checkmark$ & $\checkmark$ & $\times$  \\ \bottomrule
\end{tabular}
\end{table}

Generative large language models (LLMs) such as GPT-4~\cite{achiam2023gpt} and LLaMA-3~\cite{metaIntroducingMeta} have demonstrated exceptional capabilities of creating natural language texts across a spectrum of application domains, including chatbot~\cite{chatgpt}, coding assistant~\cite{luo2023wizardcoder,roziere2023code}, and task automation~\cite{hong2023metagpt}.
However, the increasingly large model sizes and high computational requirements of modern LLMs make it challenging to serve them cheaply and efficiently on modern cloud platforms.
In particular, most of today's LLM serving systems (e.g., Orca~\cite{yu2022orca} and vLLM~\cite{kwon2023vllm}) target {\em homogeneous} GPU clusters~\cite{miao2023towards}, where all GPUs are of the same type and have identical memory capacity and compute resources.
Due to increasing model sizes, serving LLMs using homogeneous GPUs requires an increasing number of GPUs, as shown in Table~\ref{tab:gpu_requirement}.
In addition, serving state-of-the-art LLMs used in industry requires even more resources.
Recent works have identified that it is increasingly difficult to allocate GPUs of this magnitude within a single cloud region~\cite{strati2024ml,yang2023skypilot}.

Due to advances in GPU architectural designs and the incremental deployment of them over time, modern cloud platforms increasingly consist of a mix of GPU types. 
Table~\ref{tab:sec1-cluster-heterogeneity} illustrates the {\em heterogeneous} GPU deployment in Google Compute Engine ~\cite{googlecloud2024compute}, where datacenters are equipped with various NVIDIA GPUs including H100, A100, V100, L4, and T4.
%
%
These heterogeneous GPU instances are spread across datacenters around the world and collectively offer significantly larger memory capacity and more compute resources than individual GPU types, enabling a more accessible and scalable approach to LLM serving. \camready{As Table~\ref{tab:gpu_spec} shows, eight NVIDIA L4 GPUs can offer comparable FP16 compute performance to a single NVIDIA H100 GPU, while providing greater memory capacity, lower power consumption, and a more cost-effective price point.} 
Moreover, the availability of these GPUs vary significantly across regions. We empirically find that obtaining higher quotas and securing GPU instances is much easier in some regions than others on Google Cloud Engine, and the availability of different GPUs vary a lot (see Table~\ref{tab:sec1-quota}).

\camready{Geo-distributed LLM serving with heterogeneous GPUs enables the aggregation of available GPUs from multiple regions. This approach not only enhances resource utilization but also minimizes LLM serving costs by strategically leveraging the most cost-effective GPU instances across various geographical locations.}
Similarly, there is also a trend of using volunteer consumer GPUs to address the GPU scarcity problem~\cite{ryabinin2020towards,diskin2021distributed,yuan2022decentralized}.
However, in contrast to homogeneous GPU instances, deploying LLMs on geo-distributed heterogeneous instances necessitates accommodating various GPU devices and network conditions.

\begin{table}
    \centering
    \caption{\camready{Properties of GPUs deployed in today's data centers.  We report SXM version for H100 and A100. Data collected from NVIDIA data sheet~\cite{nvidia_h100, nvidia_a100, nvidia_l4, nvidia_t4}.}
    \vspace{-0.8em}
    }
    \label{tab:gpu_spec}
    \footnotesize
    \begin{tabular}{@{}l|c|c|c|c|c@{}}
    \toprule
     & \textbf{FP16}     & \textbf{Memory} & \textbf{Bandwidth} & \textbf{Power} & \textbf{Price}         \\
                     \textbf{GPU}    & \textbf{(TFLOPs)} & \textbf{(GB)}   & \textbf{(GB/s)}    & \textbf{(W)}   & \textbf{(USD)}         \\ \midrule
    H100~\cite{nvidia_h100}                & 1979     & 80     & 3350      & 700   & 25k $\sim$40k \\
    A100~\cite{nvidia_a100}                 & 312      & 40     & 1555      & 400   & 10k $\sim$15k \\
    L4~\cite{nvidia_l4}                   & 242      & 24     & 300       & 72    & $\sim$3k      \\
    T4~\cite{nvidia_t4}                   & 65       & 16     & 300       & 70    & $\sim$1k      \\ \bottomrule
    \end{tabular}
    \vspace{-0.6em}
\end{table}
\begin{table}
\centering
\caption{
GPU quotas and deployment limits in us-east and asia-southeast during our evaluation of \sys. We measure deployment limits at two distinct time periods.
\vspace{-0.8em}
}
\label{tab:sec1-quota}
\footnotesize
\begin{tabular}{@{}l|ccc|ccc@{}}
\toprule
               & \multicolumn{3}{c|}{\textbf{Quota}} & \multicolumn{3}{c}{\textbf{Max Deployed}}  \\
\textbf{Region}         & \textbf{A100 40GB}     & \textbf{L4}     & \textbf{T4}    & \textbf{A100 40GB} & \textbf{L4} & \textbf{T4}               \\ \midrule
us-east        & 8             & 8      & 16    & 0         & 0  & 16               \\
asia-southeast & 8             & 24     & 32    & 4         & 12 & \textgreater{}20 \\ \bottomrule
\end{tabular}
\vspace{-0.8em}
\end{table}

{Prior work has introduced several systems for running machine learning computation over heterogeneous devices or geo-distributed environments. However, prior attempts either focus on long-running training workloads~\cite{park2020hetpipe,zhang2024hap, hsieh2017gaia,ryabinin2023swarm, mei2024realhf}, which cannot adapt to LLM serving scenarios with real-time inference requests, or focus on decentralized serving with volunteer computing\camready{~\cite{borzunov2022petals, borzunov2024distributed}}, which lack the global coordination necessary to efficiently use GPU and network resources in clusters.}

To efficiently serve LLMs over heterogeneous GPUs and network, we propose \sys, a distributed system for high-throughput, low-latency LLM serving. 
\sys's key idea is to formulate the execution of LLM serving over heterogeneous GPUs and network as a \textit{data flow} problem under the constraints of diverse GPU computing capabilities, memory capacities, as well as complex inter-GPU connections.
\sys leverages mixed integer linear programming to determine optimal model placement under these constraints. To accommodate heterogeneity, \sys introduces \textit{per-request pipelines}, where each request has its own independent pipeline for scheduling. This combination of flow-based formulation and per-request pipelines enables \sys to achieve high GPU utilization in heterogeneous and geo-distributed GPU clusters. We will discuss the challenges and \sys's solutions in Sec.~\ref{sec3:challenges}.
We have implemented \sys on top of vLLM~\cite{kwon2023vllm} and evaluated it on three heterogeneous clusters ranging from 24 to 42 nodes, with up to 7 different node types. The models we evaluated include \camready{LLaMA-1 30B and LLaMA-2 70B}. 
Compared to heterogeneity-aware baselines, 
\sys improves serving throughput by up to 3.3$\times$ while reducing average prompting and decoding latency by up to 66\% and 24\%. 

In summary, our contributions are:
\begin{itemize}
    \item A system for LLM serving in heterogeneous and geo-distributed GPU clusters.
    \item A max-flow formulation for LLM serving and an MILP-based algorithm to optimize model placement.
    \item Flexible flow-based per-request pipelines to maximize GPU utilization.
    \item An implementation of our techniques and an evaluation on various LLM benchmarks.
\end{itemize}
\section{Background}
\label{sec2:background}

\subsection{LLM Architecture and Serving}
\label{sec2:transformer-and-llm-serving}
Most of today's LLMs adopt a decoder-only Transformer architecture~\cite{radford2019language,brown2020language}, which begins by converting a natural language query into a sequence of tokens. 
The model then converts each token into a hidden state vector, whose size is referred to as the model's hidden size. 
A Transformer model comprises of input and output embeddings and a series of identical Transformer layers, each consisting of a self-attention and a feed-forward block. 
A self-attention block calculates the `affinity' between every pair of tokens and updates each token's hidden states based on this contextual relevance score. 
Feed-forward blocks independently modify each token's hidden state through a non-linear function.
 
Given an input sequence, a Transformer model computes the probability distribution for the next token, and samples from this probability distribution. Thus, the model applies an auto-regressive paradigm to generate the whole output sequence: 
given an {\em input prompt}, a model runs multiple iterations. 
At the first iteration, known as the {\em prompt phase}, the model processes all prompt tokens and generates the first output token. 
In subsequent iterations, known as the {\em decode phase}, the model incorporates both prompt and previously generated tokens to predict the next output token.
This iterative process stops when model produces a special end-of-sentence signal ($\langle$eos$\rangle$). Since the generation output is unpredictable, the exact number of iterations remains uncertain until the sequence is fully generated.

In addition to the unpredictable execution iterations, another feature of LLM serving is the high memory demand. The self-attention block requires all previous tokens' hidden states as inputs. To store the hidden states (known as the KV-cache) for newly generated tokens, the memory requirements keep increasing along the generation process.

To address these challenges, 
Orca~\cite{yu2022orca} presented iteration-level scheduling, which updates a batch at every iteration to avoid resource retention when a request is completed but others in the same batch need more iterations; 
vLLM~\cite{kwon2023vllm} introduces PagedAttention, managing memory for KV-cache with identical pages and allocating a new page only when a request has used up all its pages; 
multi-query~\cite{shazeer2019fast} and group-query attention~\cite{ainslie2023gqa} modifies the self-attention mechanism to reduce the size of KV-cache stored for each token.

\subsection{Distributed Model Serving}
\label{sec3:distributed-serving}
Open source LLMs now feature up to hundreds of billions of parameters, far exceeding the memory capacity of a single GPU. Consequently, serving an LLM requires multiple GPUs operating in parallel. 
Tensor Parallelism (TP)~\cite{shoeybi2019megatron} partitions the weight of each operator among GPUs, gathering the partial results on each device via an AllReduce/AllGather operation. However, TP is highly sensitive to network conditions. For every Transformer layer, it needs two communications. As a result, TP has a significant overhead in high-latency networks, and is only used among GPUs within a node. 

Conversely, Pipeline Parallelism (PP)~\cite{huang2019gpipe} assigns different operators (typically multiple layers) across GPUs to create multiple pipeline stages. It then splits inputs into microbatches, running them through the pipeline. 
PP only transmits the activation tensor at the boundary of pipeline stages. Hence, PP is much less network-sensitive. 
However, it is challenging to perfectly partition both the model and input batch, which results in pipeline bubbles. As a result, PP suffers from the device idle at pipeline bubbles, and necessitating careful schedule to be performant~\cite{agrawal2023sarathi}. 

Traditional data center setups typically assume \textit{homogeneous} clusters: uniform nodes with a uniform bandwidth. As a result, models are \textit{evenly} partitioned into pipeline stages and assigned to each devices.
{As LLMs grow in size and the latest generation GPUs remain scarce, deploying these models across heterogeneous computing devices has become a critical necessity, a challenge that previous research has not adequately addressed.}

{Deploying LLMs across geo-distributed, heterogeneous GPU clusters requires careful consideration of both hardware capabilities and network characteristics. Simple equal distribution of model layers across devices fails to maximize the potential of more powerful hardware. Moreover, the significant bandwidth differences between intra- and inter-regional network connections must inform both model placement and request scheduling decisions, particularly when infrastructure spans multiple geographical regions.}

{No prior work has focused on LLM serving on heterogeneous GPU clusters. 
The most related work is Petals~\cite{borzunov2022petals}, which performs decentralized LLM serving with volunteer computing. 
Users contribute GPUs to form a decentralized swarm and newly joined machines greedily serve the pipeline stages with least compute capacity.
For request scheduling, users greedily choose the server with lowest latency to them. Petals is effective for volunteer computing, but the lack of global coordination makes it unable to fully utilize the GPUs and network when the cluster is a known priori. 
Another relevant work is SWARM~\cite{ryabinin2023swarm}, which performs DNN training in heterogeneous clusters. It evenly partitions the model into pipeline stages. When routing a request to the next pipeline stage, it selects the replica based on real-time throughput of candidates. Our evaluation shows that such simple heuristics are not enough to achieve good performance in geo-distributed clusters with heterogeneous GPUs and network.}


\begin{figure*}[ht]
    \centering
    \begin{subfigure}[b]{0.24\linewidth}
        \centering
        \includegraphics[width=\textwidth]{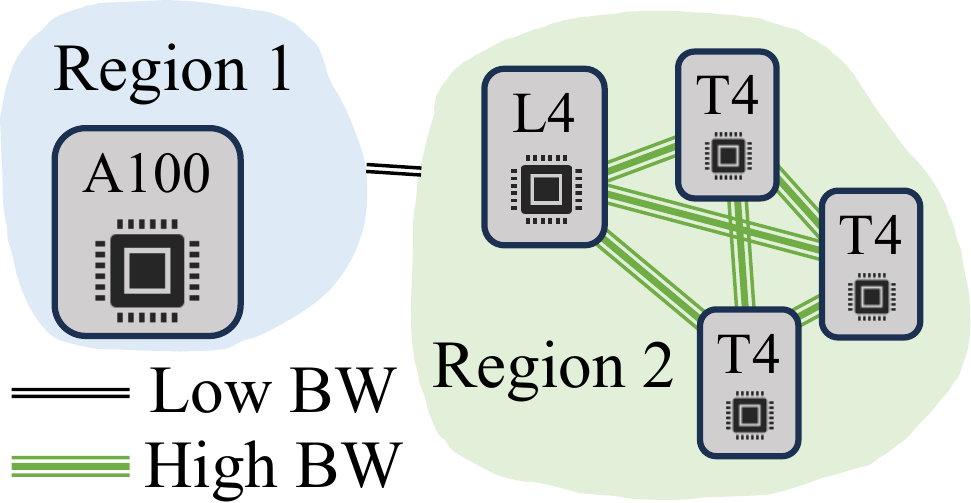}
        \caption{Five nodes in two clusters.}
        \label{fig:sec3-placement-cluster}
    \end{subfigure}
    \begin{subfigure}[b]{0.22\linewidth}
        \centering
        \includegraphics[width=\textwidth]{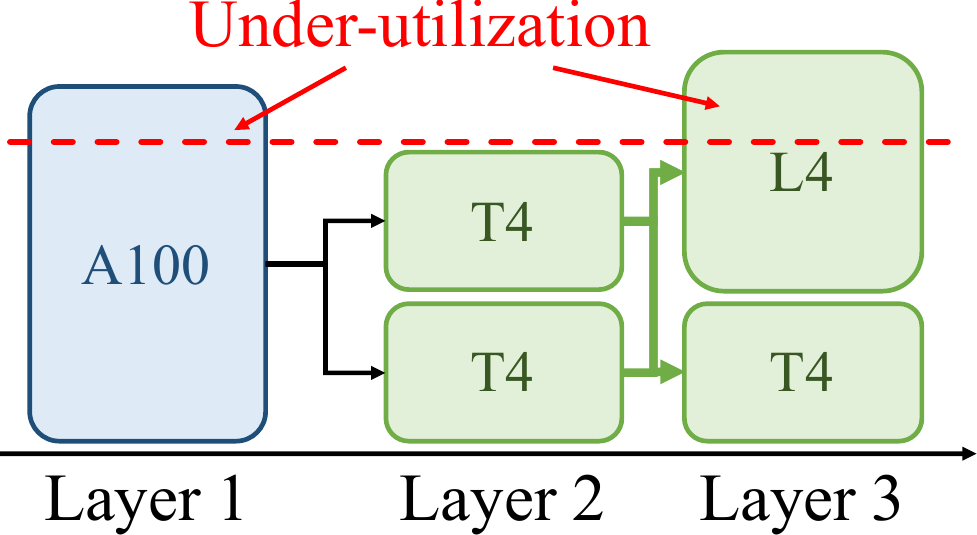}
        \caption{Uniform model partition.}
        \label{fig:sec3-placement-equal-model}
    \end{subfigure}
    \begin{subfigure}[b]{0.24\linewidth}
        \centering
        \includegraphics[width=\textwidth]{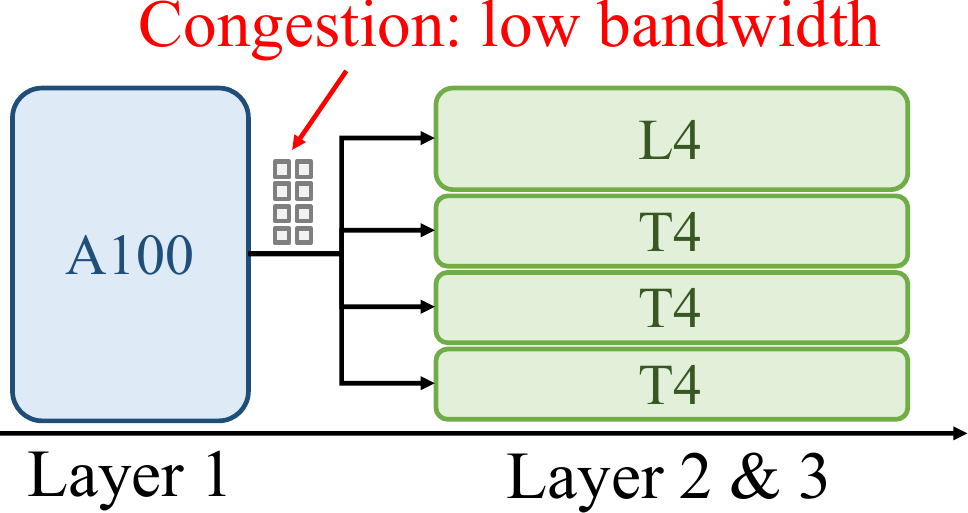}
        \caption{Balanced FLOPs.}
        \label{fig:sec3-placement-equal-compute}
    \end{subfigure}
    \begin{subfigure}[b]{0.24\linewidth}
        \centering
        \includegraphics[width=\textwidth]{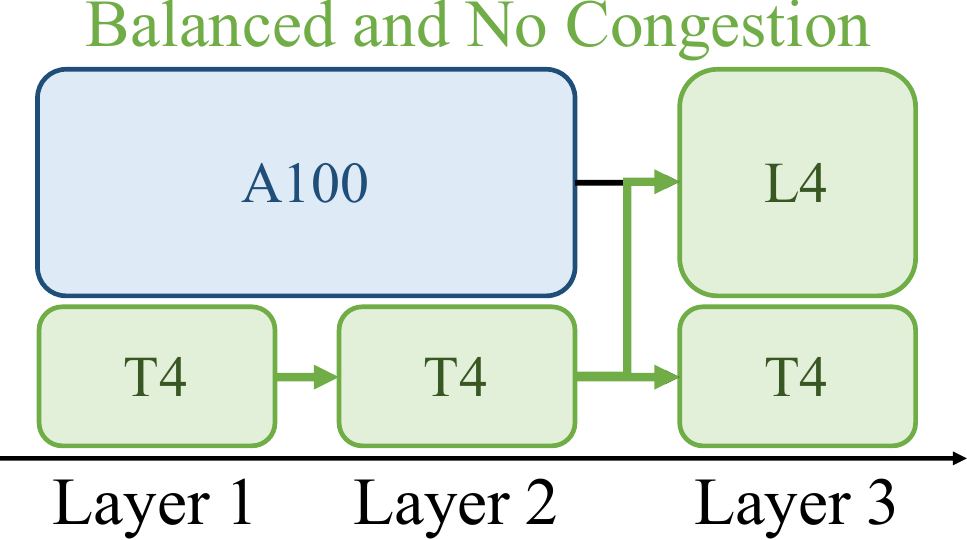}
        \caption{Network aware placement.}
        \label{fig:sec3-placement-net-aware}
    \end{subfigure}
    \vspace{-0.8em}
    \caption{Examples of sub-optimal model placement and request schedule. ~\ref{fig:sec3-placement-cluster}) all GPUs and network condition in this example. The order of compute capacity is: A100 > L4 > T4; ~\ref{fig:sec3-placement-equal-model}) Model placement by uniformly partition the model, then allocate devices by a balanced compute capacity; ~\ref{fig:sec3-placement-equal-compute}) Co-optimizing model partition and device placement to make the compute capacity more balanced; ~\ref{fig:sec3-placement-net-aware}) Co-optimizing model partition, device placement, and request scheduling in a network-aware way.
    \vspace{-0.3em}
    }
    \Description{Figure for model placement.}
    \label{fig:sec3-model-placement}
\end{figure*}
\section{Opportunities and Challenges}
\label{sec3:challenges}

Using heterogeneous and geo-distributed GPUs presents new opportunities for LLM serving. As shown in Table~\ref{tab:gpu_spec}, multiple commodity GPUs (L4, T4) can match the compute capacity of high-end GPUs (H100, A100) while offering advantages in memory capacity, energy efficiency, and cost. Furthermore, Tables~\ref{tab:sec1-quota} and~\ref{tab:network_condition} demonstrate that GPU availability varies significantly across regions, yet inter-region network conditions remain suitable for LLM serving, motivating a geo-distributed approach. However, leveraging heterogeneous and geo-distributed GPUs poses several key challenges.

\subsection{Challenge 1: Model Placement}
Due to the increasing size of LLMs, serving them on modern GPUs requires employing {\em tensor}~\cite{shoeybi2019megatron} and {\em pipeline}~\cite{park2020hetpipe, huang2019gpipe} model parallelism to partition an LLM into stages and place those stages on different GPUs, a task we term {\em model placement}. Homogeneous serving systems (e.g., Orca~\cite{yu2022orca}) partition an LLM into equal-sized stages and assign them to GPUs. This approach results in sub-optimal utilization of high-performance GPUs as it accommodates the memory and computational limitations of less powerful GPUs. Existing heterogeneity-aware serving systems (e.g., Petals~\cite{borzunov2022petals}) rely on different heuristics to partition a model into stages and assign them to GPUs. Existing heuristics do not simultaneously consider both GPU and network heterogeneity.

\paragraph{\sys's solution:} \sys exploits the flexibility of token-level scheduling in LLM serving and formulates model placement as a {\em max-flow} problem of a directed, weighted graph, whose nodes represent GPU instances and edges capture both GPU and network heterogeneity through their capacities in the max-flow problem.
\sys then uses a mixed integer linear programming (MILP) algorithm to discover highly optimized model placement strategies, which largely outperform the heuristic methods used in prior work~\cite{borzunov2022petals, ryabinin2023swarm}. Leveraging the data dependencies and homogeneity of LLM layers, \sys expresses the MILP problem with linear number of variables and constraints relative to the number of compute nodes and network connections, resulting in a tractable problem size.

\subsection{Challenge 2: Request Scheduling}

A second challenge \sys must address is {\em request scheduling}. To serve an LLM request, \sys needs to select a {\em pipeline} of GPU instances to compute all layers of the LLM. Existing systems generally employ a group of fixed pipelines and assign requests to these pipelines in a round-robin fashion. Using fixed pipelines is not flexible enough to accommodate the heterogeneous compute and network conditions and often causes under-utilization.

\paragraph{\sys's solution:} \sys introduces {\em per-request pipelines}, where each request is assigned its own pipeline. 
As a result, the total number of potential pipelines is equal to the number of paths from source to sink in the graph representation of the cluster, which offers sufficient flexibility for \sys to maximally utilize the full capacity of GPU instances and network connections between them.

\begin{figure}
    \centering
    \begin{subfigure}[b]{\linewidth}
        \centering
        \includegraphics[width=0.8\textwidth]{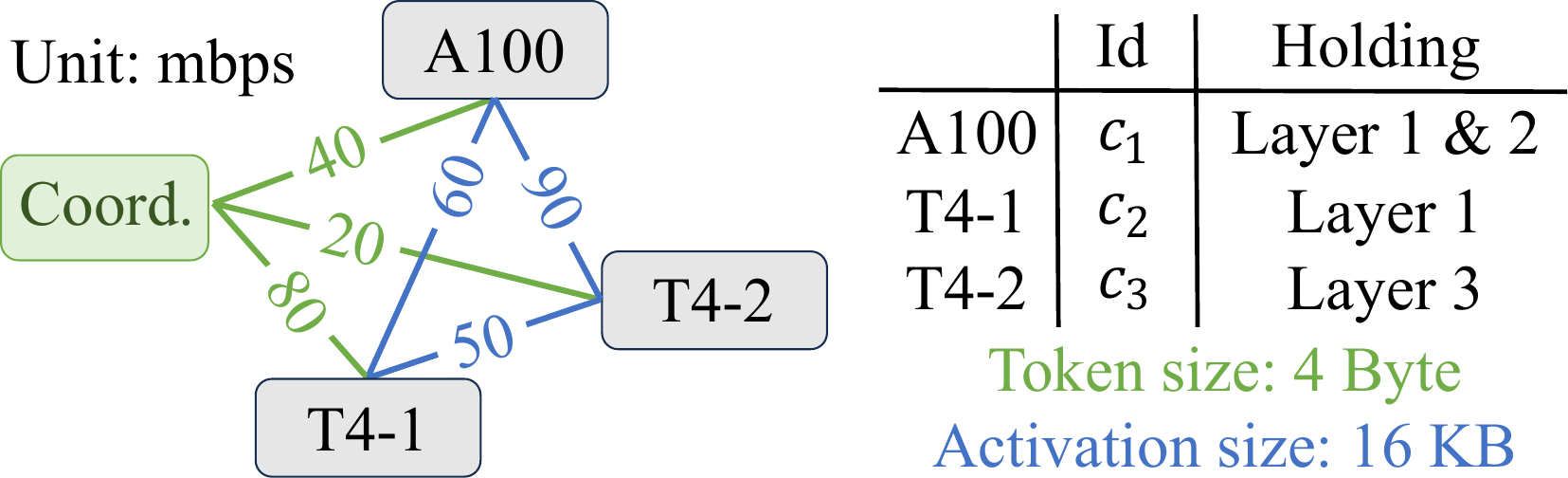}
        \caption{A 3-node cluster with model placement. Network connections between the coordinator and compute nodes transmit tokens (4 Byte) while others transmit intermediate activations (16 KB)
        }
        \label{fig:sec4-network-flow-setup}
    \end{subfigure}
    \begin{subfigure}[b]{\linewidth}
        \centering
        \includegraphics[width=0.8\textwidth]{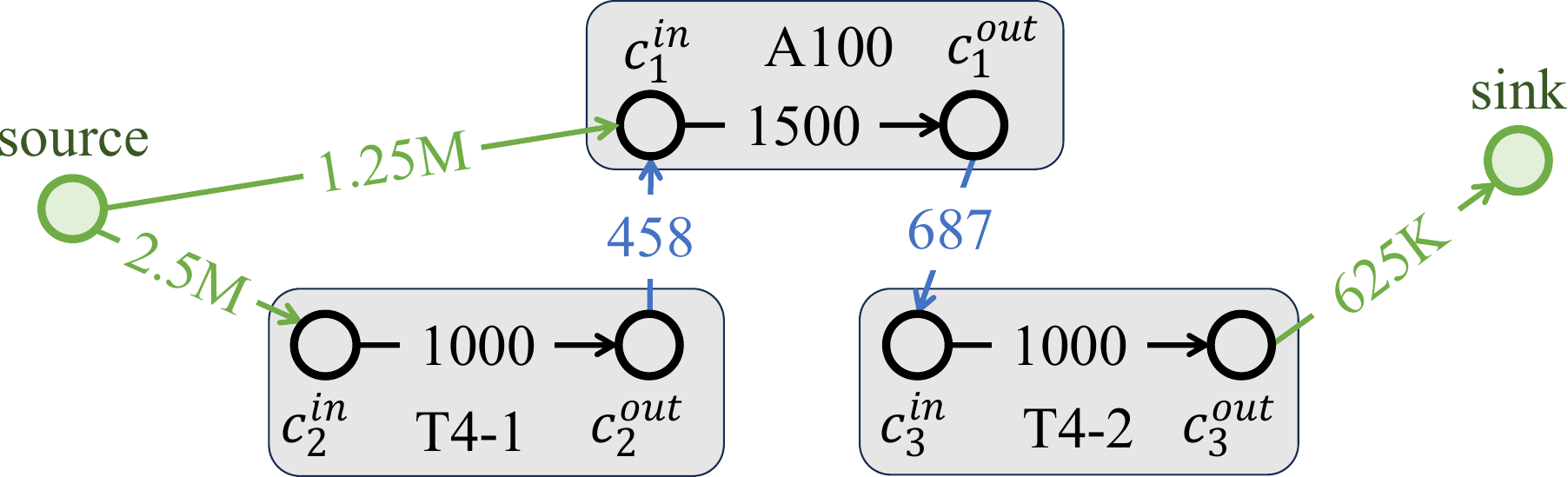}
        \caption{Graph abstraction of the cluster.}
        \label{fig:sec4-network-flow-graph}
    \end{subfigure}
    \caption{
    Graph abstraction of a 3-node cluster with given model placement. Numbers on the edges in Fig.~\ref{fig:sec4-network-flow-graph} represent their capacity, which is the number of tokens that can pass through the edges per second. Max flow between source and sink equals the max serving throughput of the cluster.
    }
    \Description{Figure for network flow.}
    \label{fig:sec4-network-flow}
\end{figure}
\section{Optimization Formulation in \sys }
\label{sec4:formulation}
This section first provides a mathematical abstraction for LLM serving systems. Based on this formulation, we model heterogeneous LLM serving as a Max-Flow problem. Finally, we apply mixed-integer linear programming (MILP) to search for a model placement strategy with the highest max flow.

\subsection{Formulation of LLM Serving}
\label{sec4:formulation-llm-serving}


A cluster to serve LLMs generally contains one coordinator node $h$ and a group of compute nodes $\mathcal{C}$. Each compute node $c_i\in \mathcal{C}$ has a compute capacity and GPU VRAM size. \camready{Compute nodes with multiple GPUs can be abstracted as a single logical node, aggregating GPUs' combined computational capacity and GPU VRAM resources.} {Throughput and latency of network connections between nodes in the cluster are also given}.
Based on the cluster information, LLM serving requires finding a placement of model layers to compute nodes to maximize a \textit{serving performance metric} that will be defined below.
The model placement function $\Psi: \mathcal{C} \mapsto \mathcal{P}(\mathcal{M})$ takes as input a compute node and returns a (usually continuous) subset of the model $\mathcal{M}$. \camready{Here, we assume pipeline parallelism for inter-node parallelization and tensor parallelism for intra-node parallelization across GPUs, as tensor parallelism demands extensive communication and may be constrained by network.}
One widely used metric~\cite{borzunov2022petals, ryabinin2023swarm} to assess a model placement is the performance of the pipeline stage with the lowest compute capacity $\min_i \sum_j \text{capacity}(j) \cdot 1_{i\in \Psi(j)}$,
{where capacity$(j)$ is the compute capacity of the $j^{th}$ node.} As we will show below, only considering compute capacity yields sub-optimal model placements for heterogeneous clusters.

Based on model placement, LLM serving requires a request scheduling strategy that can efficiently serve requests in the cluster. The request scheduling strategy $\phi: \mathcal{R} \mapsto \mathcal{C}^k$ inputs a request and outputs a sequence of compute nodes that form a complete pipeline for executing all layers of the LLM.

\subsection{{Necessity of Joint Optimization}}

Before diving into our Max-Flow formulation of heterogeneous LLM serving, we first use an example to show why we need to co-optimize model partition, device placement, and request scheduling as a Max-Flow problem. In this example, the clusters are shown in Fig.~\ref{fig:sec3-placement-cluster}. There are two regions with a low bandwidth between them. Region 1 has a powerful A100 GPU, while Region 2 has a less powerful L4 GPU and three T4 GPUs, but has a high bandwidth within the region. The pairwise bandwidths are independent. 
{If we follow the common approach, which statically partitions the model and then assigns devices to each partition, the placement plan will be as Fig.~\ref{fig:sec3-placement-equal-model}.} 
In this plan, although the last pipeline stage has a T4 and an L4 GPU, its throughput is bound by the previous stage's output throughput, which only has 2 T4 GPUs. 
This indicates a necessity to co-optimize the pipeline partition plan and placement of pipeline stages.

However, even with a perfectly balanced compute capacity at each pipeline stage, as shown in Fig.~\ref{fig:sec3-placement-equal-compute}, the solution can still be sub-optimal. 
In this solution, it assigns the powerful A100 to individually serve some layers, while other GPUs run in a data parallel manner for the rest of the layers. 
However, communications from one pipeline stage to another become a bottleneck. For every request, its intermediate state is sent from Region 1 to Region 2 via low bandwidth. This eventually creates congestion on the A100's send side. 
Instead, Fig.~\ref{fig:sec3-placement-net-aware} assigns two T4 GPUs running in parallel with the A100. This divides the workload between the A100 GPU and the two T4 GPUs, reducing communication on the slow link.

\subsection{Heterogeneous LLM Serving as Max-Flow}
\label{sec4:formulation-max-flow}

To optimize model placement, \sys needs a way to determine the max serving throughput of different model placements. To achieve this, we transform a cluster of compute nodes with assigned model layers into a directed graph with edge capacity. The edge capacity denotes the number of tokens compute nodes and network connections can process/transmit per second. 
Max flow between source and sink vertices in the graph, which represent the coordinator node, gives us the max serving throughput of the cluster with the current model placement. The following shows the formal construction.


\begin{table}
\centering
\caption{Variables used in MILP. 
\vspace{-0.8em}
}
\label{tab:ilp-variables}
\scalebox{0.8}{
\begin{tabular}{@{}ccc|l@{}}
\toprule
Symbol & Type & Num. & Description \\ \midrule
$s_i$    &   int   &   $O(|\mathcal{C}|)$   &     index of $c_i$'s first layer        \\
  $b_i^{j}$  &   binary   &   $O(|\mathcal{C}|)$    &   whether $c_i$ holds $j$ layers          \\
$f_{i,j}$   &   real   &   $O(|\mathcal{E}|)$   &   flow from $c_i$ to $c_j$          \\
$d_{i,j}$ & binary & $O(|\mathcal{E}|)$ & whether $(c_i, c_j)$ is valid \\
$cond^{1}_{i,j}$ & binary & $O(|\mathcal{E}|)$ & aux. variable in constraint-4 \\
$cond^{2}_{i,j}$ & binary & $O(|\mathcal{E}|)$ & aux. variable in constraint-4 \\
\bottomrule
\end{tabular}
}
\vspace{-0.5em}
\end{table}

{For a given cluster with coordinator node $h$, a set of compute nodes $\mathcal{C}$, and a model placement $\Psi$, we can transform entities in the cluster into elements of its graph abstraction as follows. An example of such a graph abstraction of a cluster with \textit{given model placement} is shown in Fig.~\ref{fig:sec4-network-flow}}

\paragraph{Compute and coordinator nodes.} For each compute node $c_i \in \mathcal{C}$, we represent it with two connected vertices in the graph. We name the two vertices $c^{in}_i$ and $c^{out}_i$. The capacity of the directed edge $(c^{in}_i, c^{out}_i)$ represents the max number of tokens this node can process in one second. It is the minimum of the node's compute and network throughput. 
\sys performs a one-time profiling to measure the throughput of all compute nodes. 
For the coordinator node, we represent it as source and sink vertices in the graph.


\paragraph{Network connections.} In a given cluster, a node may communicate with any other nodes, creating $O(|\mathcal{C}|^2)$ possible directed network connections between different nodes. However, only a subset of those connections are valid based on the model placement as described below. A {\em valid connection} should satisfy one of the following three criteria: (1) the connection is from coordinator node $h$ to compute node $c_i$ and $c_i$ holds the first layer of the model; (2) the connection is from a compute node $c_j$ to coordinator node $h$ and $c_j$ holds the last layer of the model; (3) the connection is from one compute node $c_i$ to another compute node $c_j$ and $c_j$ holds model layers immediately needed after inference on $c_i$. For the first and second case, we represent the connection with directed edge $(source, c^{in}_i)$ and $(c^{out}_j, sink)$ respectively, with capacity equal to the connection bandwidth divided by the transmission size of a token (a few bytes). For the third case, we represent the connection with a directed edge $(c_i^{out}, c_j^{in})$, and the capacity equals the connection bandwidth divided by the transmission size of an activation (tens of kilobytes). \camready{\sys performs a one-time profiling and uses the average bandwidth as the connection bandwidth.} The capacity of the edges models the throughput constraint imposed by the speed of network connection between different nodes. We denote the full set of possible network connections by $\mathcal{E}$.

After constructing the equivalent graph abstraction of a cluster, we run the preflow-push algorithm~\cite{cheriyan1989analysis} to get the max flow between source and sink node. One unit of flow here represents one token that can pass through a compute node or network connection in one second. Therefore, the max flow gives us the max possible serving throughput of the cluster with current model placement.

\begin{table}
\centering
\caption{Constraints used in MILP. 
\vspace{-0.8em}
}
\label{tab:ilp-constraints}
\scalebox{0.8}{
\begin{tabular}{@{}l|c|l@{}}
\toprule
Group                            & Num.  & Constraint \\ \midrule
\multirow{2}{*}{Model placement} & $O(|\mathcal{C}|)$     &       $\sum_{j=1}^{k} b_{i}^{j} = 1$     \\
                                 & $O(|\mathcal{C}|)$     &   $0\leq s_i < L$ and $e_i \leq L$         \\
                                 \midrule
Flow conservation                & $O(|\mathcal{C}|)$     &       $\sum_u f_{u,i} = \sum_v f_{i,v}$     \\ \midrule
Infer. throughput                  & $O(|\mathcal{C}|)$     &   $\sum_u f_{ui} \leq \sum_{j=1}^k b_{i}^j \cdot T_j$         \\ \midrule
\multirow{5}{*}{Connection validity}     & $O(|\mathcal{C}|)$   &      $s_i \leq L \cdot (1 - d_{source,i})$      \\
                                 & $O(|\mathcal{C}|)$   &     $L \cdot d_{i,sink} \leq e_i$       \\
                                 & $O(|\mathcal{E}|)$ &       $(L + 1)(1 - cond^{1}_{i,j}) \geq s_j - e_i$     \\
                                 & $O(|\mathcal{E}|)$ &       $e_j - e_i \geq 1 - (L + 1)(1 - cond^{2}_{i,j})$     \\
                                 & $O(|\mathcal{E}|)$  &      $d_{i,j} \leq 0.5 * cond^{1}_{i,j} + 0.5 * cond^{2}_{i,j}$      \\ \midrule
Trans. throughput                  & $O(|\mathcal{E}|)$     &      $f_{i,j} \leq d_{i,j} \cdot S_{i,j}$      \\  \bottomrule
\end{tabular}
}
\end{table}

\subsection{Optimal Model Placement with MILP}
\label{sec4:formulation-milp}
The previous section presented an approach for obtaining the max serving throughput of a cluster \textit{for a given model placement}.
In this section, we introduce a mixed-integer linear programming (MILP)-based method to find a model placement that maximizes the max flow, thus maximizing serving throughput. 
{The MILP formulation has a linear number of variables and constraints with respect to the number of compute nodes and network connections. The key challenges addressed include (1) formulation of system-level constraints as linear number of conditions to satisfy, (2) expression of these conditions with linear number of variables, and (3) linearization of each condition using auxiliary variables, specifically, each constraint is expressed as at most three linear constraints with the help of at most two auxiliary variables.}
An overview of the variables and constraints is shown in Table~\ref{tab:ilp-variables} and \ref{tab:ilp-constraints}.

\paragraph{Node variables.} To represent the model placement on each compute node, we introduce two groups of variables in our MILP formulation. Suppose that the model has a total of $L$ layers and each compute node holds a continuous subset of the model. For each compute node $c_i$, we introduce an integer variable $s_i$ to represent the first layer $c_i$ holds. Suppose compute node $c_i$ can hold at most $k$ layers on its GPU, we further introduce $k$ binary variables $b_{i}^{1}, b_{i}^{2}, ..., b_{i}^{k}$ to indicate the number of layers node $c_i$ holds ($b_{i}^{j}=1$ if $c_i$ holds $j$ layers). We choose to express model placement with $k$ binary variables (instead of one integer for the number of layers) because this formulation facilitates the expression of inference throughput constraints as discussed below.
{The \textit{end layer index} of $c_i$ can be expressed as $e_i = s_i + \sum_{j=1}^{k} j \cdot b_{i}^{j}$. Therefore, $c_i$ holds layers in range $[s_i, e_i)$.}

\begin{figure*}[ht]
    \centering
    \includegraphics[width=0.8\linewidth]{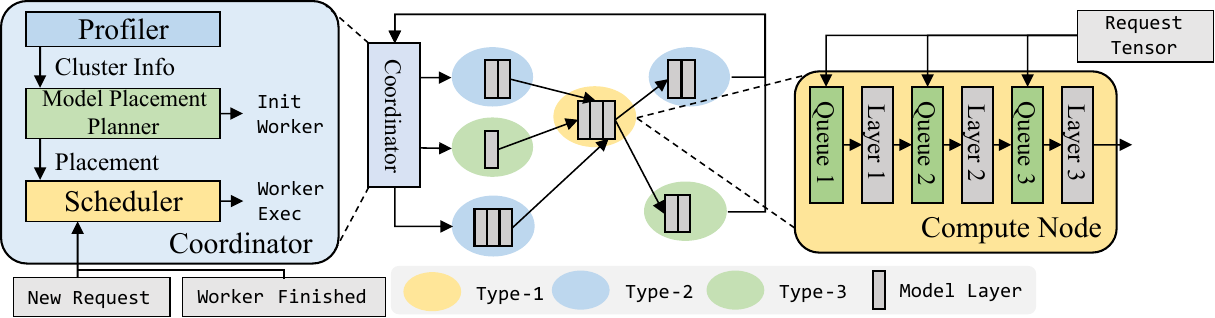}
    \caption{\textbf{\sys overview.} In \sys, the coordinator plans model placement as described in Sec.~\ref{sec4:formulation-milp}. We only need to run model placement once for each cluster. When a new request arrives, the coordinator node runs \sys scheduler to assign it a per-request pipeline and sends it to the first node in the pipeline. {Each compute node in the pipeline performs inference on the request on the layers it is responsible for and sends the (output for the) {request} to the next node in the pipeline. When the last node in the pipeline finishes performing inference on its layers, it will send the output token for the request to the coordinator (Worker Finished). The coordinator schedules generation of the next token for the request using the same pipeline.}
    }
    \vspace{-0.2em}
    \Description{Figure for system overview.}
    \label{fig:sec5-overview}
\end{figure*}

\paragraph{Connection variables.} We introduce two groups of variables to constrain the number of inference requests that can go through each network connection. For network connection between compute node $c_i$ and $c_j$, we introduce a real variable $f_{i,j}$ to denote the amount of flow from $c_i^{out}$ to $c_j^{in}$ in the graph abstraction. We further introduce a binary variable $d_{i,j}$ to denote whether the network connection is valid (as defined in Sec.~\ref{sec4:formulation-max-flow}). The constraints we introduce below will use $d_{i,j}$ to ensure that requests can only be transmitted through valid connections. For network connections between coordinator node and compute nodes, we similarly introduce two variables similar to above, but replace $i$/$j$ with source/sink.

\paragraph{Constraint-1: model placement.} To ensure that the model placement found by the MILP solver is valid, we need the following two constraints for each compute node $c_i$. First, $c_i$ should have only one valid model placement, meaning that $\sum_{j=1}^{k} b_{i}^{j} = 1$. Moreover, the first and last layer $c_i$ holds must be within the range of $L$ layers, meaning that $0\leq s_i < L$ and $e_i \leq L$. ($c_i$ holds layers in range $[s_i, e_i)$)

\paragraph{Constraint-2: flow conservation.} For each compute node $c_i$, the sum of flow that goes in to $c_i^{in}$ must be equal to that goes out of $c_i^{out}$ because of flow conservation. This constraint can be expressed as $\sum_u f_{u,i} = \sum_v f_{i,v}$, where $u$ and $v$ enumerate through all nodes except $i$.

\paragraph{Constraint-3: inference throughput.} For compute node $c_i$, the amount of flow that passes through $(c_i^{in}, c_i^{out})$ should not exceed its maximum inference throughput. We can impose this constraint with $\sum_u f_{ui} \leq \sum_{j=1}^k b_{i}^j \cdot T_j$. {Here, $T_j$ is a constant that represents the maximum number of tokens node $c_i$ can process in one second when holding $j$ layers, which is obtained through a one-time profiling process.}

\paragraph{Constraint-4: connection validity.} We need to determine the validity of network connections to know if requests can be transmitted through them. For a network connection from the coordinator node to the compute node $c_i$, it is valid only if $c_i$ holds the first layer of the model. To express this constraint with MILP, we need to linearize it into the following form: $s_i \leq L \cdot (1 - d_{source,i})$. Similarly, for network connection from compute node $c_i$ to coordinator, we constrain its validity with $L \cdot d_{i,sink} \leq e_i$. For network connection from compute node $c_i$ to $c_j$, its validity $d_{i,j}$ is determined by whether $s_j \leq e_i < e_j$ holds. To linearize this condition, we need to introduce two binary auxiliary variables $cond^{1}_{i,j}$ and $cond^{2}_{i,j}$. $cond^{1}_{i,j}$ takes value 1 only if $s_j \leq e_i$, which can be linearized as $(L + 1)(1 - cond^{1}_{i,j}) \geq s_j - e_i$. $cond^{2}_{i,j}$ takes value 1 only if $e_i < e_j$, which can be linearized as $e_j - e_i \geq 1 - (L + 1)(1 - cond^{2}_{i,j})$. The network connection is valid only if both binary auxiliary variables are true, which can be expressed as $d_{i,j} \leq 0.5 * cond^{1}_{i,j} + 0.5 * cond^{2}_{i,j}$.

We remark that if $s_j < e_i < e_j$, then requests coming from $c_i$ will only infer layers $[e_i, e_j)$ on $c_j$. We call this {\em partial inference}. If partial inference is not allowed, then the connection validity constraints can be simplified to $d_{i,j}=1$ only if $e_i = s_j$, which linearizes to two constraints $L \cdot d_{i,j} \leq L + s_j - e_i$ and $L \cdot d_{i,j} \leq L - s_j + e_i$.

\paragraph{Constraint-5: transmission throughput.} We only allow flow to pass through valid network connections, and the flow should not be larger than the connection's maximum transmission throughput. {To enforce this constraint, we add $f_{i,j} \leq d_{i,j} \cdot S_{i,j}$ as a constraint into the MILP problem.} $S_{i,j}$ is the maximum number of tokens that can be transmitted through the network connection, which can be calculated via profiling and using methods mentioned in Sec.~\ref{sec4:formulation-max-flow}.

\paragraph{Optimization target.} The MILP problem aims to find a model placement that satisfies all constraints and yields the highest max flow for the cluster. This optimization target can be expressed as maximizing the sum of flow from source, i.e. maximizing $\sum_i f_{source,i}$.

\paragraph{MILP solution orchestration.} After the MILP solver finds a solution that satisfies all constraints, we can orchestrate it into a model placement plan and construct the graph abstraction of the cluster. For compute node $c_i$, $s_i$ and $e_i$ give us the model layers $c_i$ should load into its GPU.

\subsection{Analyzing and Speeding up MILP}
\label{sec4:milp-analysis}

As Table~\ref{tab:ilp-variables} and \ref{tab:ilp-constraints} show, the number of variables and constraints in the MILP problem scales linearly with the number of compute nodes and network connections. {For large clusters with more than 40 nodes, it may still take hours before the MILP solver gives a reasonably good solution.}
To expedite the MILP solving process for large clusters, we introduce three optimizations. First, we prune some of the slow network connections in the cluster. {Evaluation in Sec.~\ref{sec6:ablation} shows that this effectively reduces the problem size without sacrificing much performance}.
Second, we hint the MILP solver with solutions found by heuristic methods. {Since the problem has an exponential solution space, the MILP solver can only cover a small portion within a limited solving time budget. Using solutions from heuristic methods as starting points for the MILP problem expedites the optimization process, especially for large clusters. 
Sec.~\ref{sec6:ablation} shows the necessity of starting from heuristic solutions for large clusters.}
{Finally, we notice that the max serving throughput of a cluster is always bounded by the sum of compute throughput of all compute nodes averaged by the total number of layers. The MILP solver uses this as an early stop criterion and stops when it finds a solution that is very close to this upper bound.}
\camready{We remark that, for further scaling of \sys to hundreds or even thousands of nodes, one viable approach is to first partition the nodes into multiple smaller clusters using heuristics and then apply \sys independently.}

\subsection{\camready{Replacing MILP with LP or Heuristics?}}

A common approach for speeding up MILP problems is to relax them to a linear program (LP) by relaxing the integer variables to be linear variables and obtaining a valid solution to the original problem via methods such as rounding the resulting linear variables. We remark that this approach is not viable for the MILP problem above.
This is because the resulting solution from the LP cannot be easily converted to a valid solution of the original problem. The variables for model placement ($s_i$ and $b_i^j$) decide the edge validity variables $d_{i,j}$, which in turn decides the flow variables $f_{i,j}$. Rounding the non-integral values of model placement variables in the relaxed solution may invalidate some or all network connections and thus drastically changing the max flow.

\camready{Our preliminary exploration also indicates that using simpler heuristics cannot guarantee good performance across various cluster setups, because of the exponential solution space of this problem. Sec.~\ref{sec6:model-placement} shows that the model placement found by \sys with MILP is much better than that of the heuristic baselines.}


\begin{figure}
    \centering
    \includegraphics[width=\linewidth]{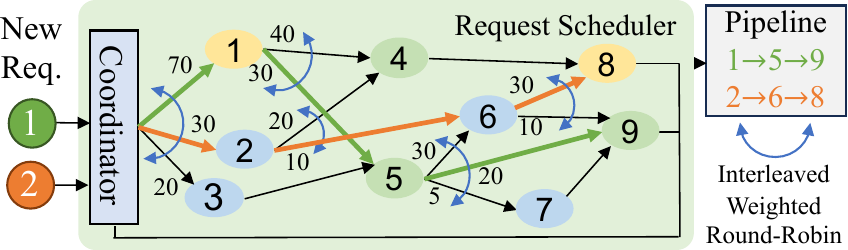}
    \caption{Topology graph of a cluster, where each vertex is a compute node, and each edge is a valid network connection. Numbers over edges represent the flow over the network connection in the max flow solution. The pipelines used to schedule requests 1 and 2 are shown on the right. 
    \vspace{-0.8em}
    }
    \label{fig:sec5-iwrr}
    \Description{Figure for IWRR.}
\end{figure}

\section{\sys Runtime}
\label{sec5:runtime}

This section discusses the runtime scheduling of requests in \sys. {When the coordinator node receives a new request, it runs \sys's request scheduler to assign the request a {\em per-request pipeline}, which we will introduce in Sec.~\ref{sec5:scheduler-design}.}
Then the coordinator node sends the request to the first compute node in the pipeline. {When a compute node receives a requests, it performs inference on the request using the layers it is responsible for in that pipeline} 
and sends the request to the following compute node. Fig.~\ref{fig:sec5-overview} shows the overview of \sys. 

\subsection{Scheduler Design: Per-Request Pipelines}
\label{sec5:scheduler-design}
To infer a request in the cluster, the scheduler needs to assign a {\em pipeline} for the request. The pipeline contains a sequence of stages, where each stage specifies a compute node and the layers to infer on the compute node. A valid pipeline must infer each layer of the model exactly-once and in correct order when running the stages sequentially. \camready{Existing works~\cite{borzunov2022petals, jiang2023hexgen} use fixed pipelines, in which each pipeline contains a disjoint set of machines, and assign requests to those pipelines. In \sys, instead of using fixed-pipelines, we propose a \textit{per-request pipeline} assignment approach, wherein each request will have its own pipeline and the pipelines may intersect with each other.} The total number of possible pipelines equals the number of possible paths from source to sink in the graph abstraction of the cluster. The abundant number of pipelines allows the scheduling to better fit the capacity of the compute nodes and network connections. Our Max-Flow formulation enables us to create the per-request pipelines.

{The \sys request scheduler performs scheduling based on the cluster's \textit{topology graph} (Fig.~\ref{fig:sec5-iwrr}). In the topology graph, vertices correspond to the nodes in the cluster. Directed edges correspond to the valid network connections (under the model placement found by solving the MILP in Sec.~\ref{sec4:formulation}). We bind an interleaved weighted round-robin (IWRR)~\cite{tabatabaee2021interleaved} scheduler to each vertex. The IWRR scheduler takes as input a list of candidates and their weights. To schedule a request, it selects a candidate with frequency proportional to its weight. For each vertex $u$, the IWRR scheduler's candidates contain all vertices $v$ such that directed edge $(u, v)$ exists (i.e. $(u, v)$ represents a valid network connection). The weight of $v$ equals the flow over the network connection of $(u, v)$ in the max flow solution. Using IWRR allows us to schedule requests following the max flow without creating bursts.}

{\sys's request scheduler runs on the coordinator node. When a new request arrives, it first uses the IWRR scheduler of the vertex representing the coordinator node to determine the compute node $c_1$ for the first pipeline stage. Then, it uses the IWRR scheduler of the vertex representing $c_1$ to determine the compute node $c_2$ for the second stage, and repeats this process until a valid pipeline is established. Fig.~\ref{fig:sec5-iwrr} shows the scheduling of two requests. After setting the request's pipeline, the coordinator node sends the request to the first compute node in the pipeline to begin inference.} \camready{During inference, \sys adopts a \textit{dynamic batching} strategy where each node includes all requests received during the processing of the previous batch to form a new one. This best-effort batching occurs without additional waiting periods.}

\subsection{KV-Cache Estimation}
\label{sec5:kv-estimation}
When serving LLMs, each GPU has a limited amount of VRAM to store the KV-cache of requests during inference. If the requests running concurrently on the GPU require more KV-cache than this limit, the execution engine has to offload some requests to main memory, which significantly harms throughput. However, we do not know exactly how much KV-cache each request will use because the length of the output is unknown before inference finishes. Therefore, in the scheduler we maintain an estimation of KV-cache usage of all compute nodes using average output length, and mask out compute nodes that exceed the high water mark when running IWRR. We can schedule more requests to the compute nodes only after some requests currently running on those nodes have finished. This mask ensures that we do not oversubscribe the GPU's KV-cache.


\begin{figure}
    \centering
    \begin{subfigure}[b]{0.48\linewidth}
        \centering
        \includegraphics[width=\textwidth]{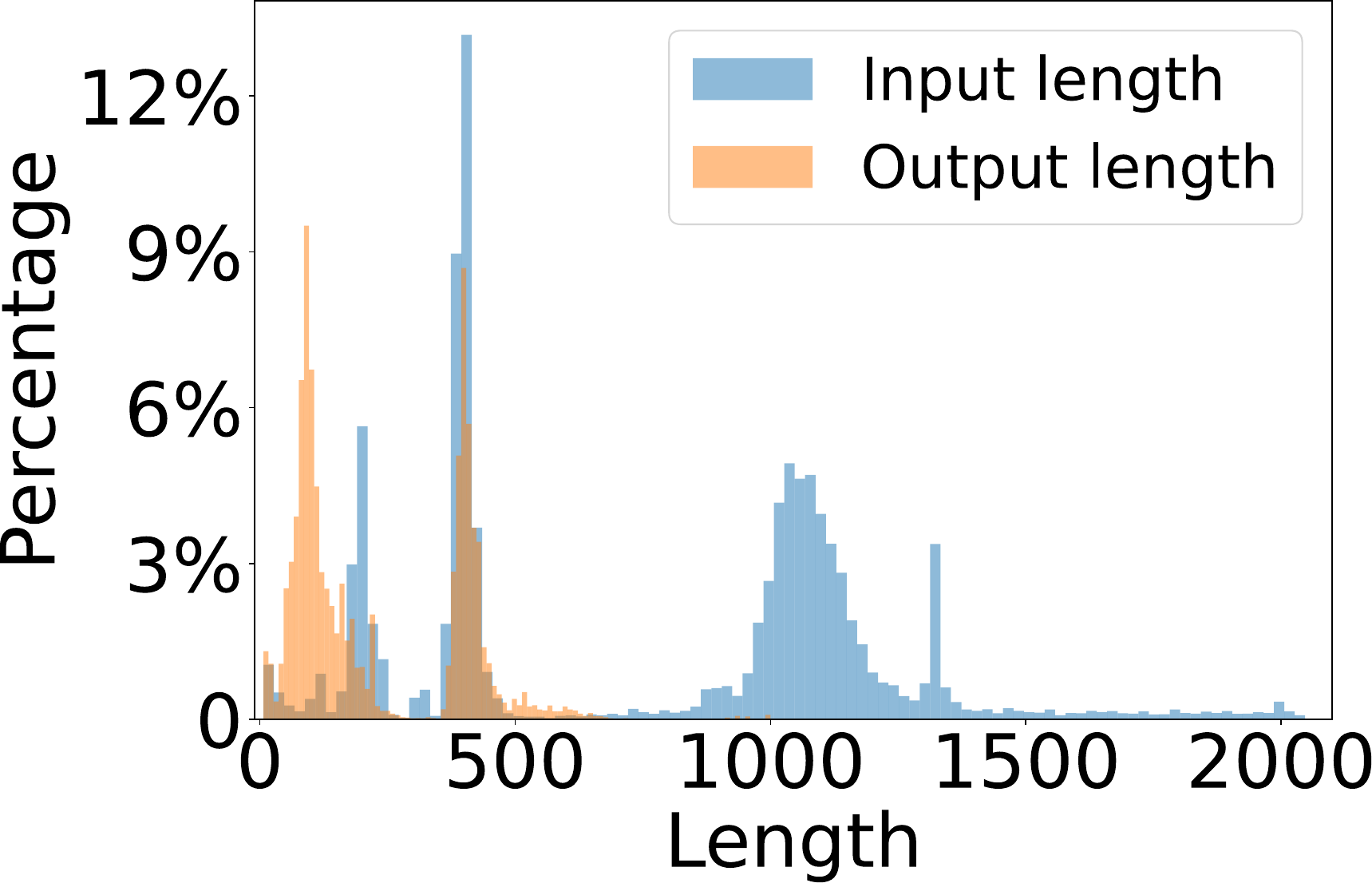}
        \caption{Length distribution.}
        \label{fig:sec6-azure-conv-length}
    \end{subfigure}
    \hfill
    \begin{subfigure}[b]{0.48\linewidth}
        \centering
        \includegraphics[width=\textwidth]{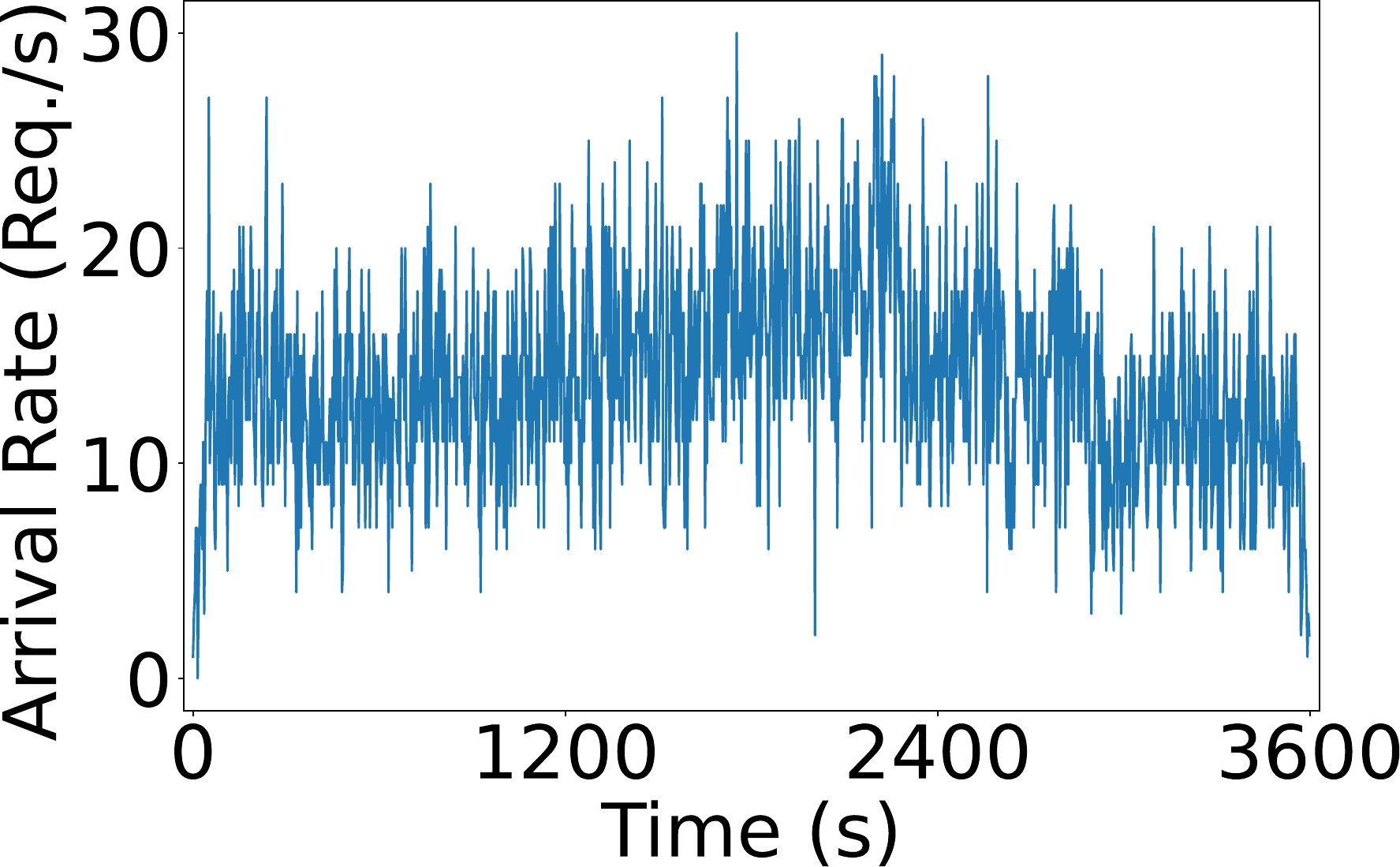}
        \caption{Arrival rate.}
        \label{fig:sec6-azure-conv-arrival}
    \end{subfigure}
    \vspace{-0.8em}
    \caption{Statistics of Azure Conversation dataset. 
    \vspace{-0.8em}
    }
    \label{fig:sec6-azure}
    \Description{Figure for Azure Conversion dataset.}
\end{figure}
\section{Evaluation}
\label{sec6:evaluation}

\begin{table}
\centering
\caption{\camready{Network bandwidth between machines in asia-east, us-central, europe-west and australia-southeast. Measured with iperf3 on Google Compute Engine.}
\vspace{-0.4em}
}
\label{tab:network_condition}
\footnotesize
\begin{tabular}{@{}c|cccc@{}}
\toprule
\multirow{2}{*}{Receiver} & \multicolumn{4}{c}{Sender}                                    \\
                          & asia-east2-a & us-central1-f & eu-west3-c & au-se1-c \\ \midrule
asia-east2-a                & /            & 123 Mbps      & 67 Mbps        & 175 Mbps      \\
us-central1-f               & 122 Mbps     & /             & 204 Mbps       & 123 Mbps      \\
eu-west3-c              & 61 Mbps      & 196 Mbps      & /              & 54 Mbps       \\
au-se1-c             & 159 Mbps     & 118 Mbps      & 63 Mbps        & /             \\ \bottomrule
\end{tabular}
\end{table}
\begin{figure*}[t]
    \centering
    \begin{subfigure}{0.24\linewidth}
    \centering
    \includegraphics[width=\linewidth]{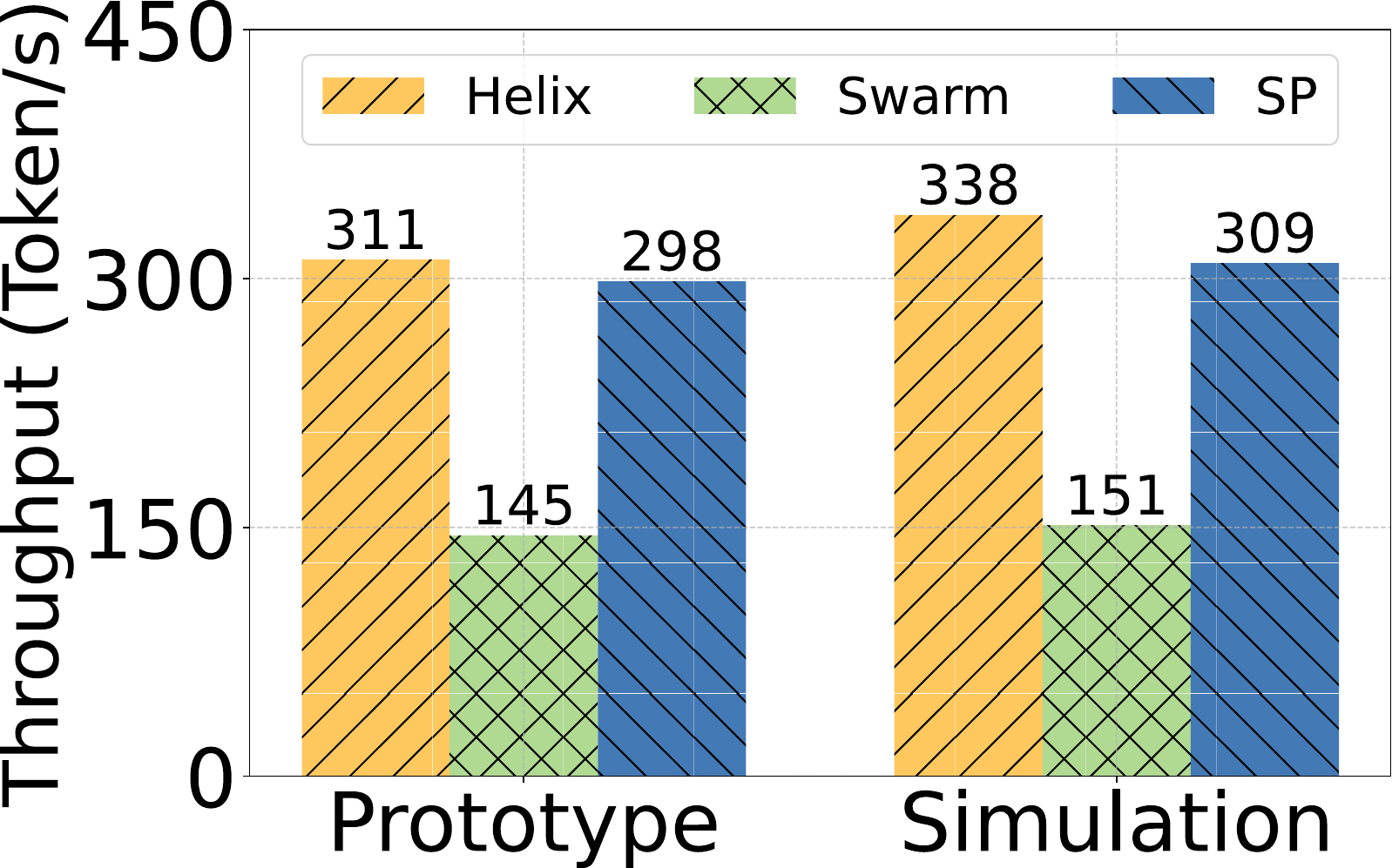}
    \caption{LLaMA-30B - Offline}
    \label{fig:sec6-single-30b-offline-throughput}
    \end{subfigure}
    \begin{subfigure}{0.24\linewidth}
    \centering
    \includegraphics[width=\linewidth]{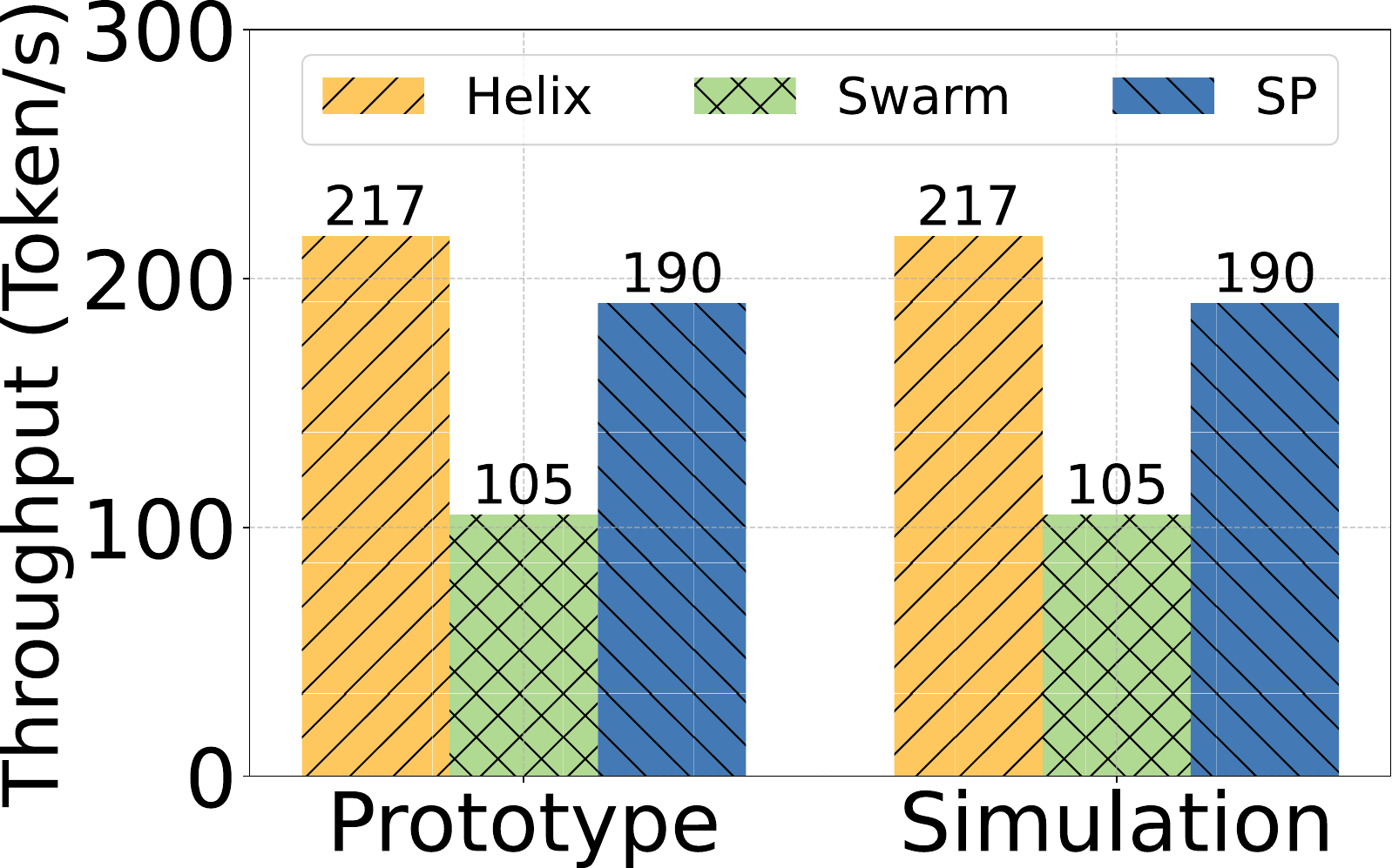}
    \caption{LLaMA-30B - Online}
    \label{fig:sec6-single-30b-online-throughput}
    \end{subfigure}
    \begin{subfigure}{0.24\linewidth}
    \centering
    \includegraphics[width=\linewidth]{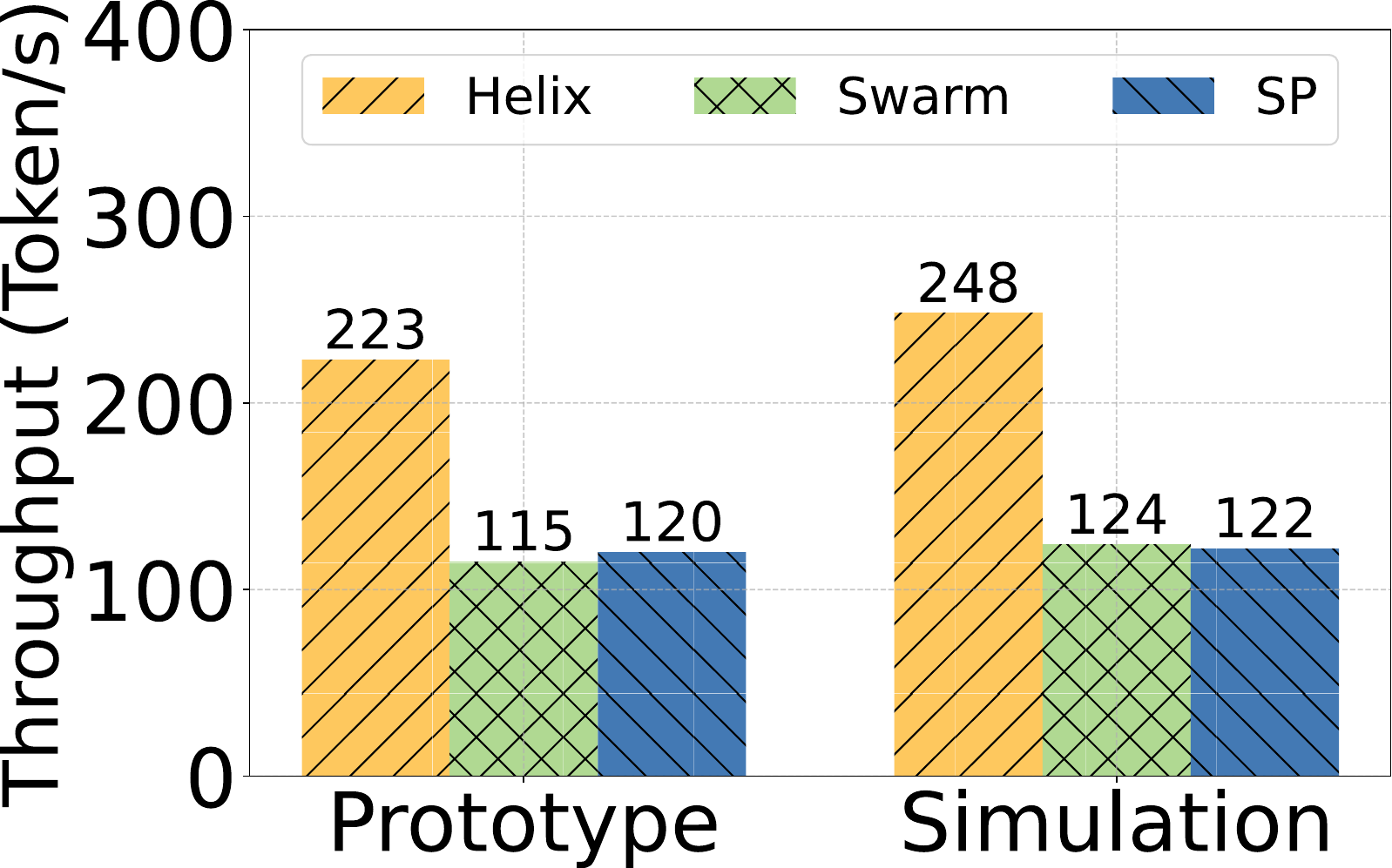}
    \caption{LLaMA-70B - Offline}
    \label{fig:sec6-single-70b-offline-throughput}
    \end{subfigure}
    \begin{subfigure}{0.24\linewidth}
    \centering
    \includegraphics[width=\linewidth]{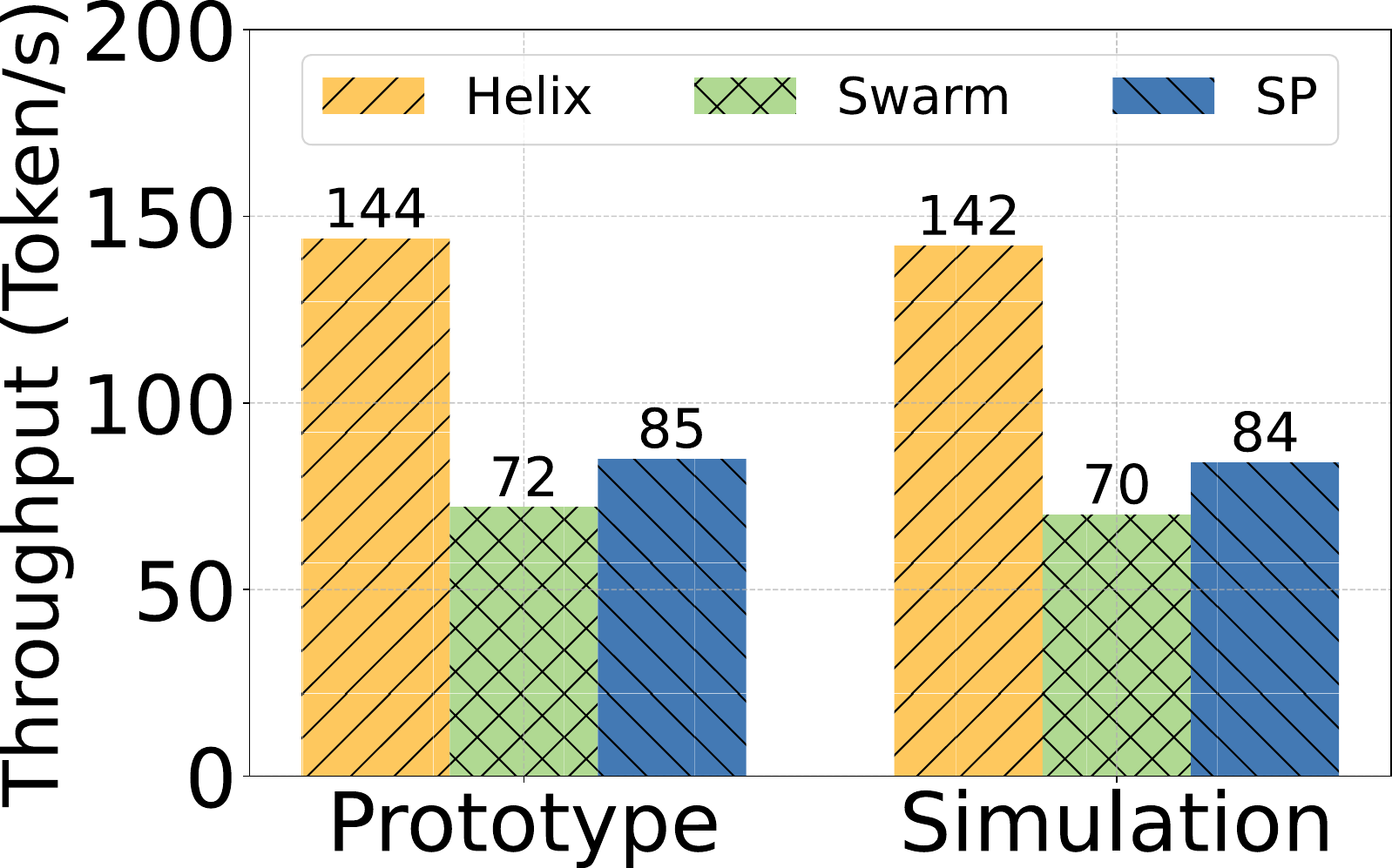}
    \caption{LLaMA-70B - Online}
    \label{fig:sec6-single-70b-online-throughput}
    \end{subfigure}
    \begin{subfigure}{0.24\linewidth}
    \centering
    \includegraphics[width=\linewidth]{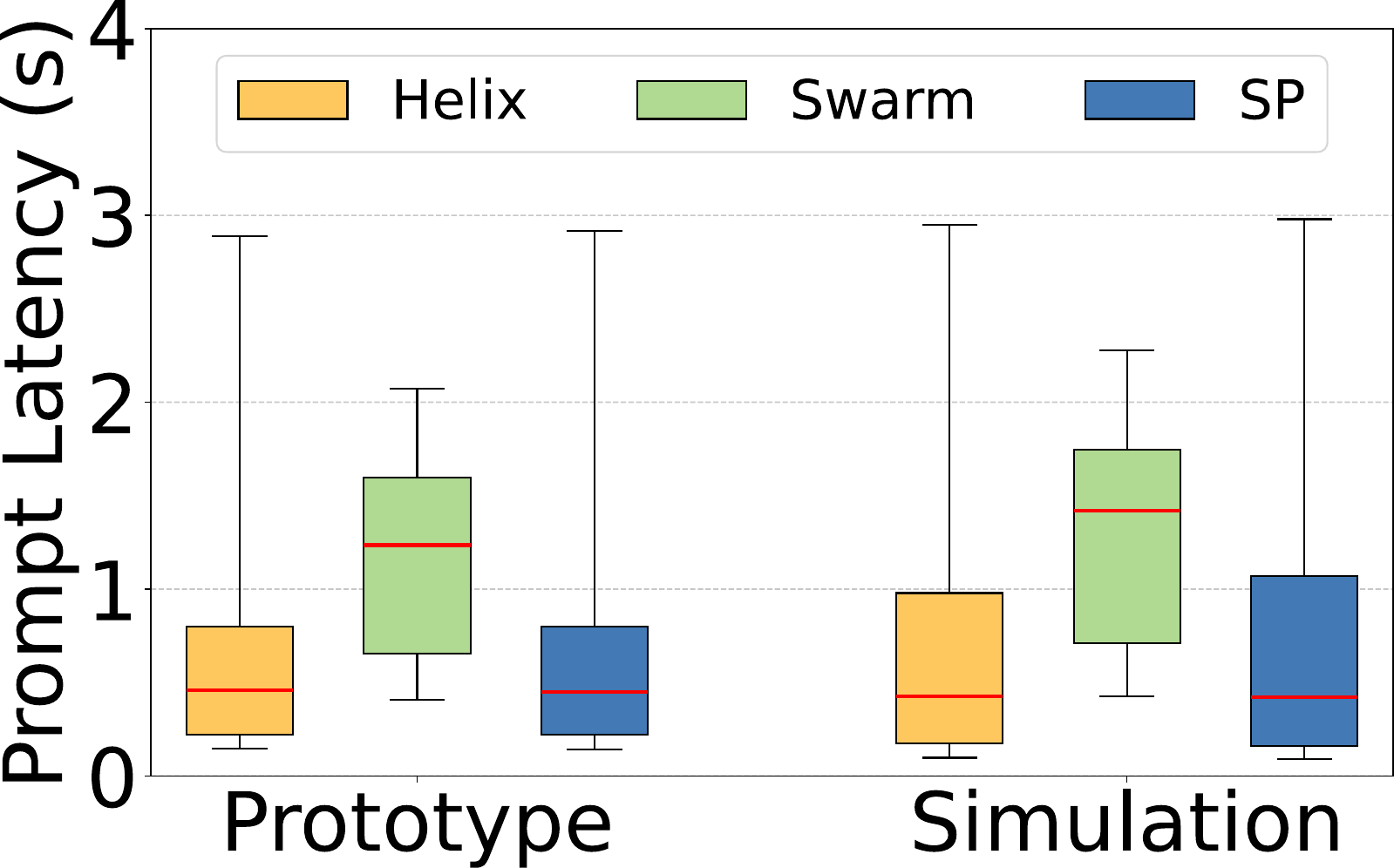}
    \caption{LLaMA-30B - Prompt}
    \label{fig:sec6-single-30b-online-prompt}
    \end{subfigure}
    \begin{subfigure}{0.24\linewidth}
    \centering
    \includegraphics[width=\linewidth]{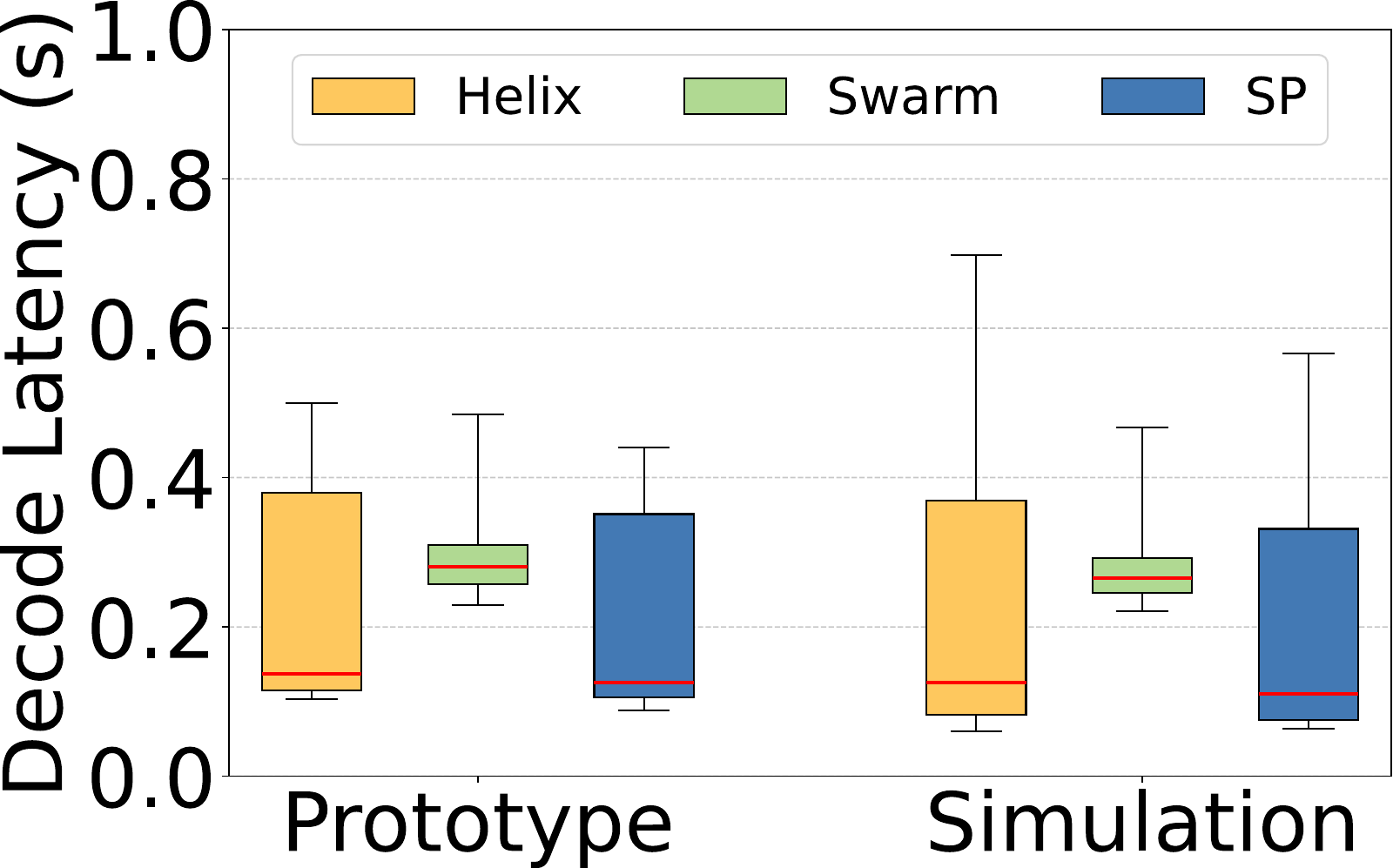}
    \caption{LLaMA-30B - Decode}
    \label{fig:sec6-single-30b-online-decode}
    \end{subfigure}
    \begin{subfigure}{0.24\linewidth}
    \centering
    \includegraphics[width=\linewidth]{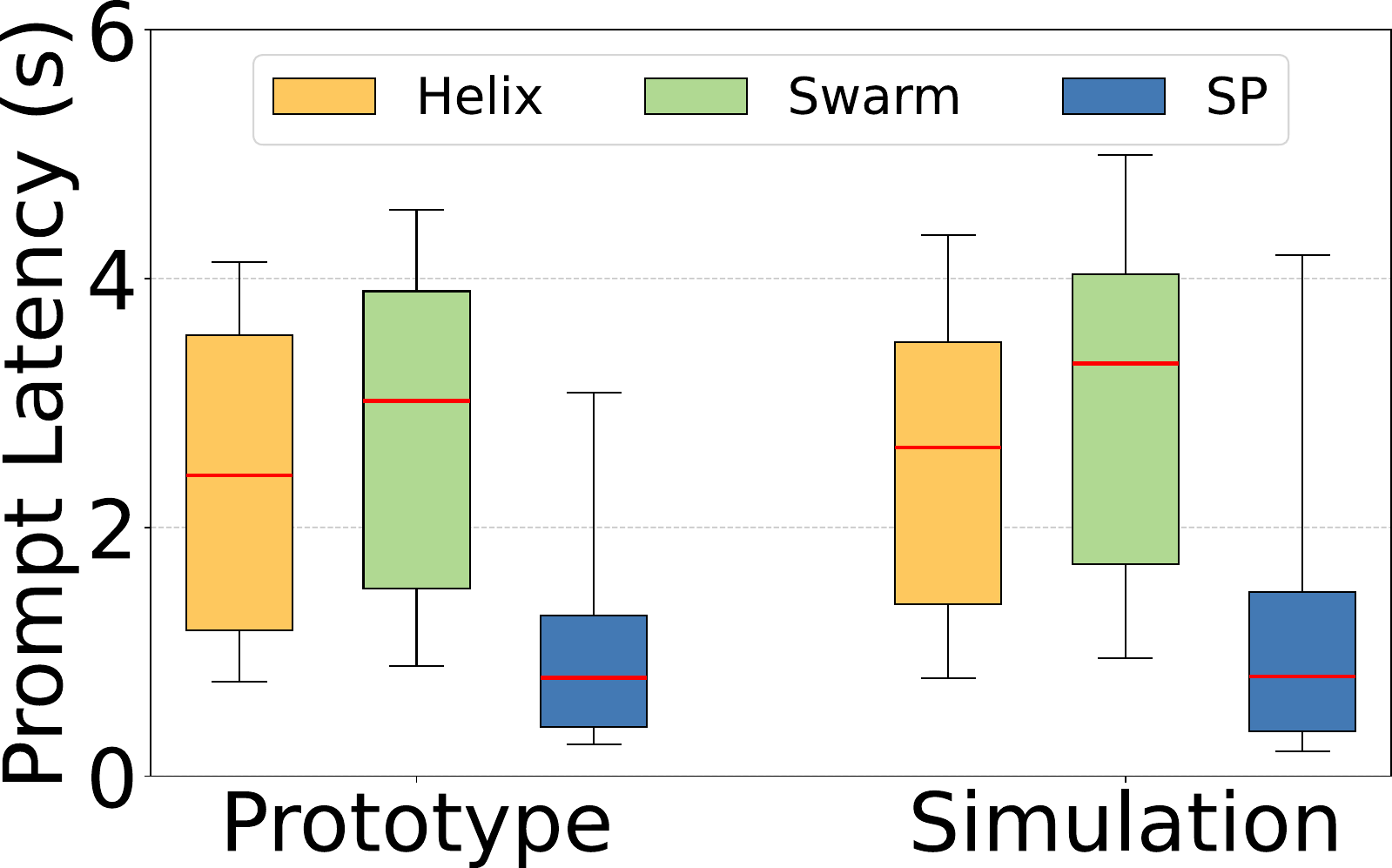}
    \caption{LLaMA-70B - Prompt}
    \label{fig:sec6-single-70b-online-prompt}
    \end{subfigure}
    \begin{subfigure}{0.24\linewidth}
    \centering
    \includegraphics[width=\linewidth]{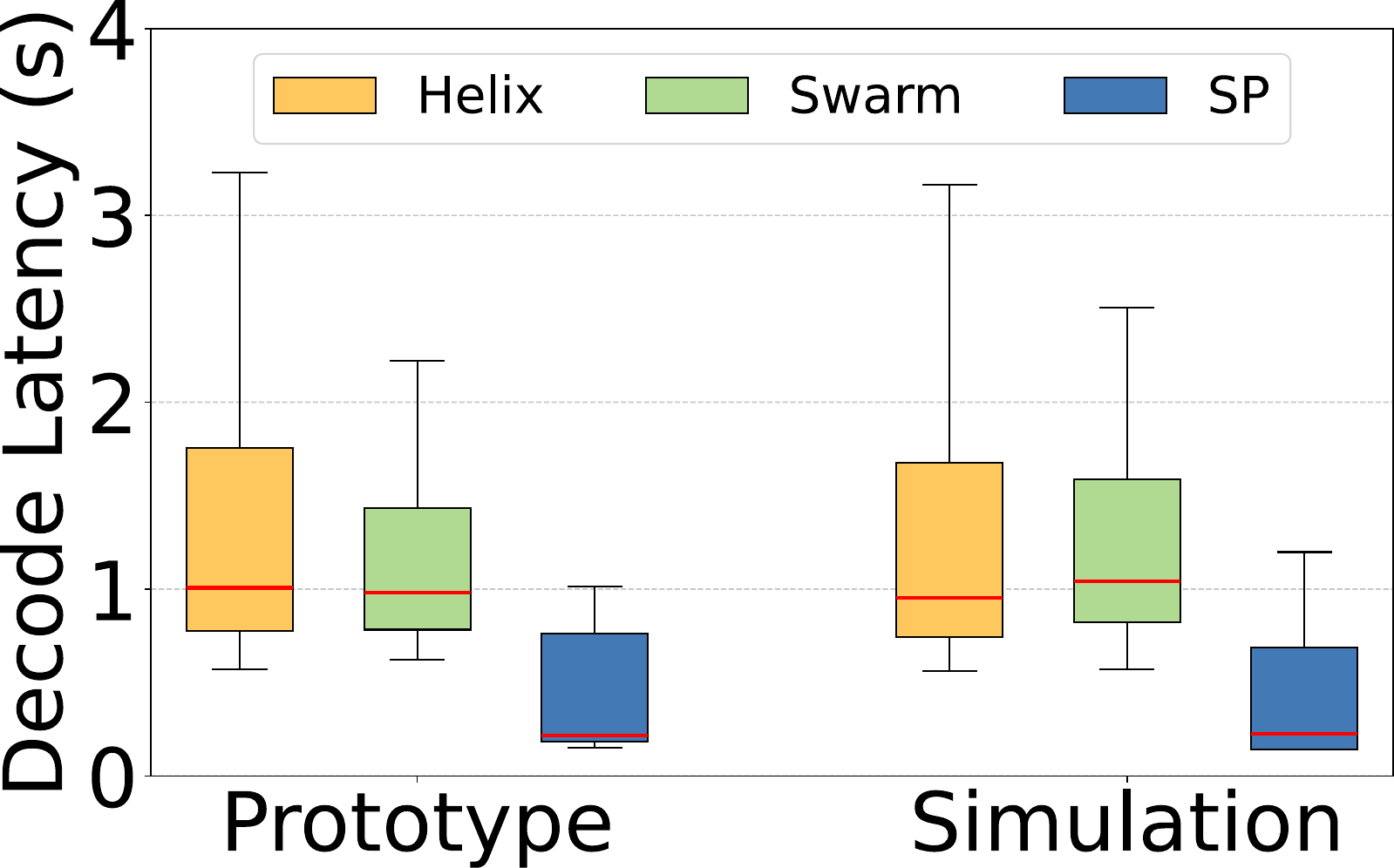}
    \caption{LLaMA-70B - Decode}
    \label{fig:sec6-single-70b-online-decode}
    \end{subfigure}
    \vspace{-0.8em}
    \caption{\textbf{Single cluster results:} \textbf{(a - d)} decode throughput, \textbf{(e - h)} prompt and decode latency for online serving. The box shows the 25 and 75 percentile. The whisker shows the 5 and 95 percentile. The red line shows median.}
\label{fig:sec6-single-cluster}
\Description{Figure for single cluster results.}
\end{figure*}

In this section, we aim to answer the following questions.
\begin{itemize}[noitemsep,nolistsep]
    \item {Can \sys provide higher throughput for heterogeneous GPUs in a single cluster? (Sec.~\ref{sec6:single-cluster})}
    \item {Does \sys sacrifice latency for throughput? (Sec.~\ref{sec6:single-cluster})}
    \item {Can \sys provide higher throughput for heterogeneous GPUs in geo-distributed clusters? (Sec.~\ref{sec6:distributed-clusters})}
    \item {Can \sys provide consistently high throughput when the degree of GPU heterogeneity increases? (Sec.~\ref{sec6:cluster_heterogeneity})}
    \item Why does \sys achieve better performance compared to existing systems (Sec.~\ref{sec6:model-placement} and \ref{sec6:request_scheduling})?
\end{itemize}


\subsection{Implementation}
\label{sec6:implementation-details}

\paragraph{\sys prototype.} {We implemented a multi-replica pipeline parallel system with 1.5k LoC in Python and 1.7k LoC in C++. 
For model execution, we adopt the latest release of vLLM~\cite{kwon2023vllm} (0.4.0post1) to avoid re-implementing basic LLM inference optimizations. We implemented a \textit{unified page pool} atop vLLM to support partial inference. We use ZeroMQ~\cite{zmq} for inter-node communication and Gurobi~\cite{gurobi} as MILP solver.}

\paragraph{Simulator.} {We also implemented a simulator for distributed LLM inference with 14k LoC in Python. It supports simulation for heterogeneous GPUs and network conditions. It gives us the flexibility to explore more diverse settings for network, GPU heterogeneity and cluster scale. Sec.~\ref{sec6:single-cluster} evaluates the fidelity of the simulator and shows that the  simulation errors are less than 5\% for all metrics.}

\begin{figure*}[t]
    \centering
    \begin{subfigure}{0.19\linewidth}
    \centering
    \includegraphics[width=\linewidth]{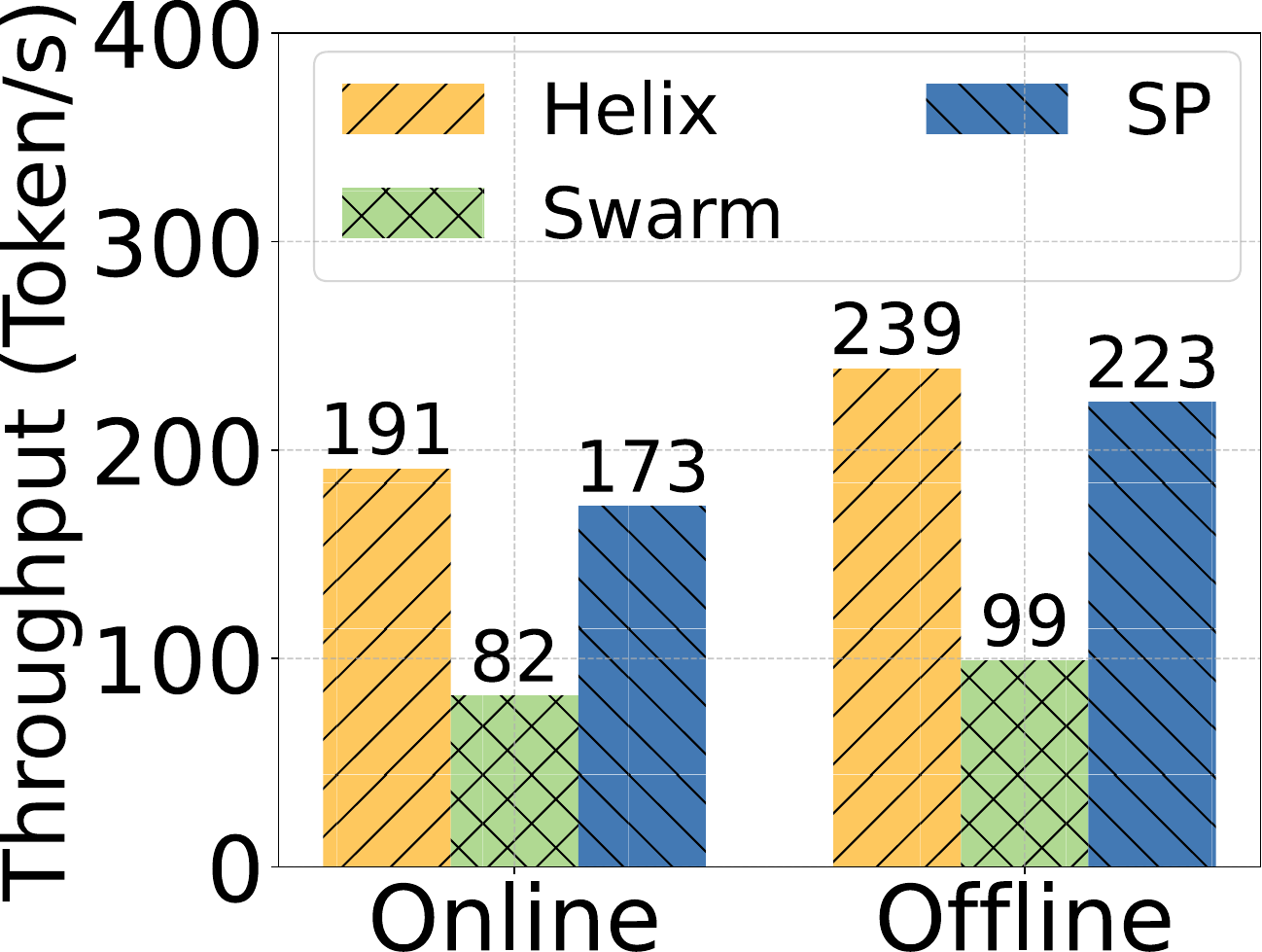}
    \caption{LLaMA-30B}
    \label{fig:sec6-distributed-30b-throughput}
    \end{subfigure}
    \begin{subfigure}{0.19\linewidth}
    \centering
    \includegraphics[width=\linewidth]{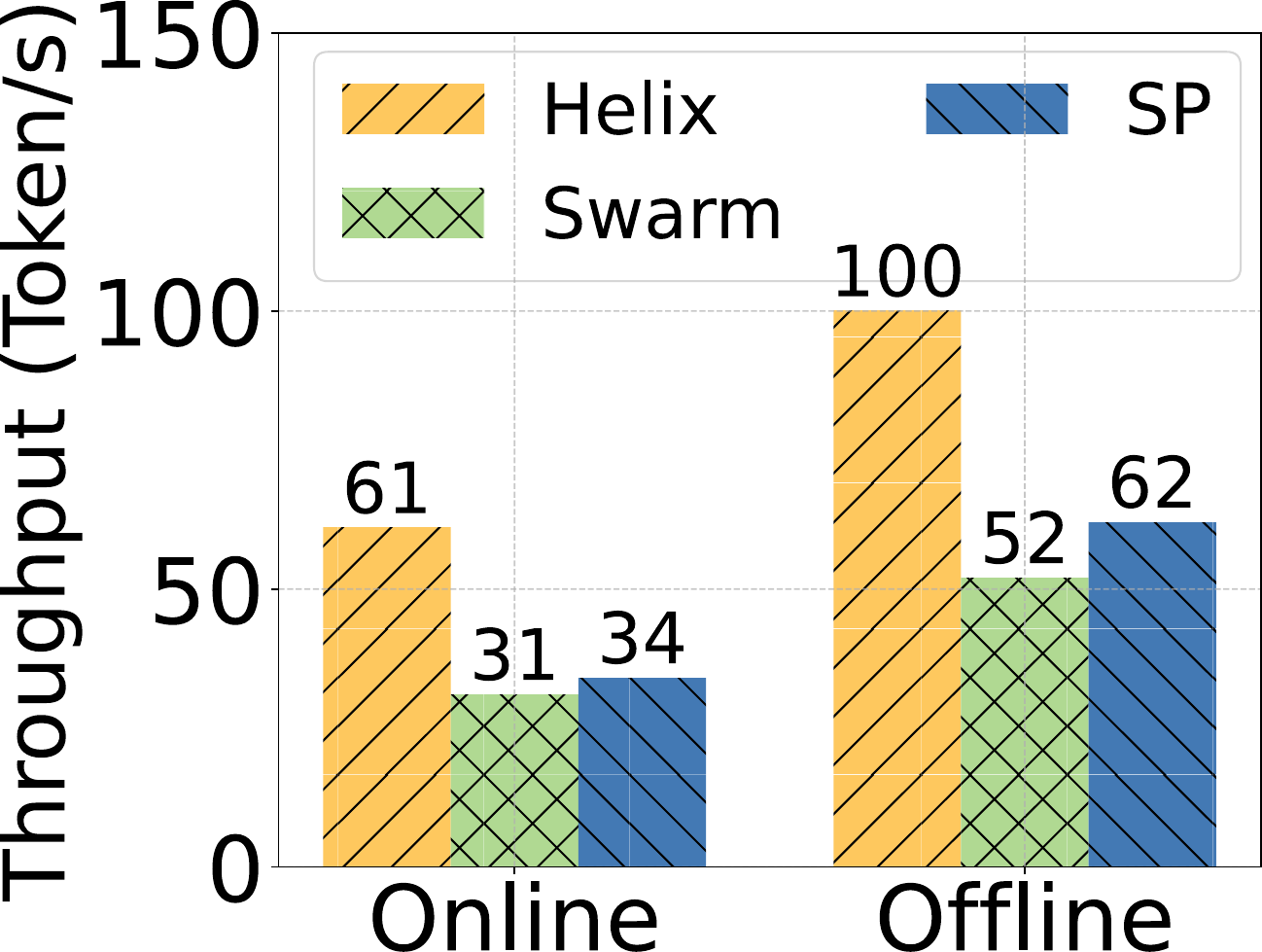}
    \caption{LLaMA-70B}
    \label{fig:sec6-distributed-70b-throughput}
    \end{subfigure}
    \begin{subfigure}{0.14\linewidth}
    \centering
    \includegraphics[width=\linewidth]{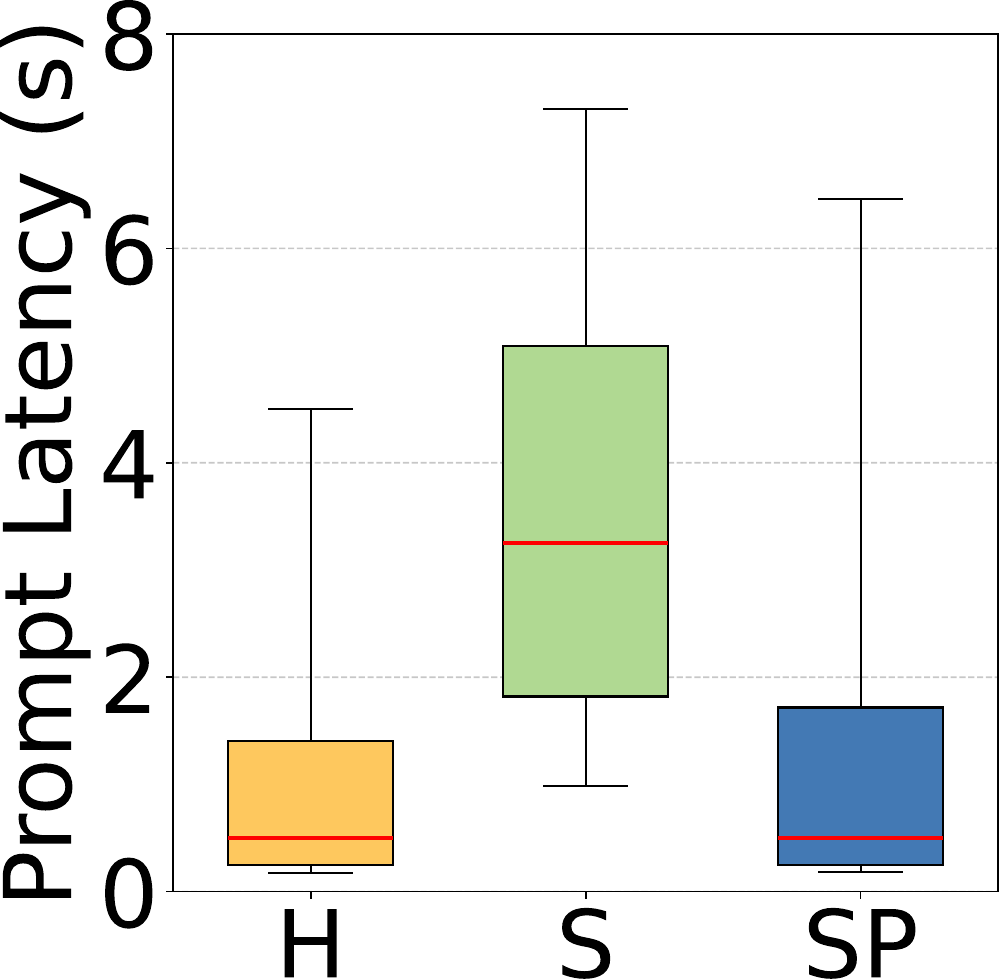}
    \caption{LLaMA-30B}
    \label{fig:sec6-distributed-30b-prompt-latency}
    \end{subfigure}
    \begin{subfigure}{0.15\linewidth}
    \centering
    \includegraphics[width=\linewidth]{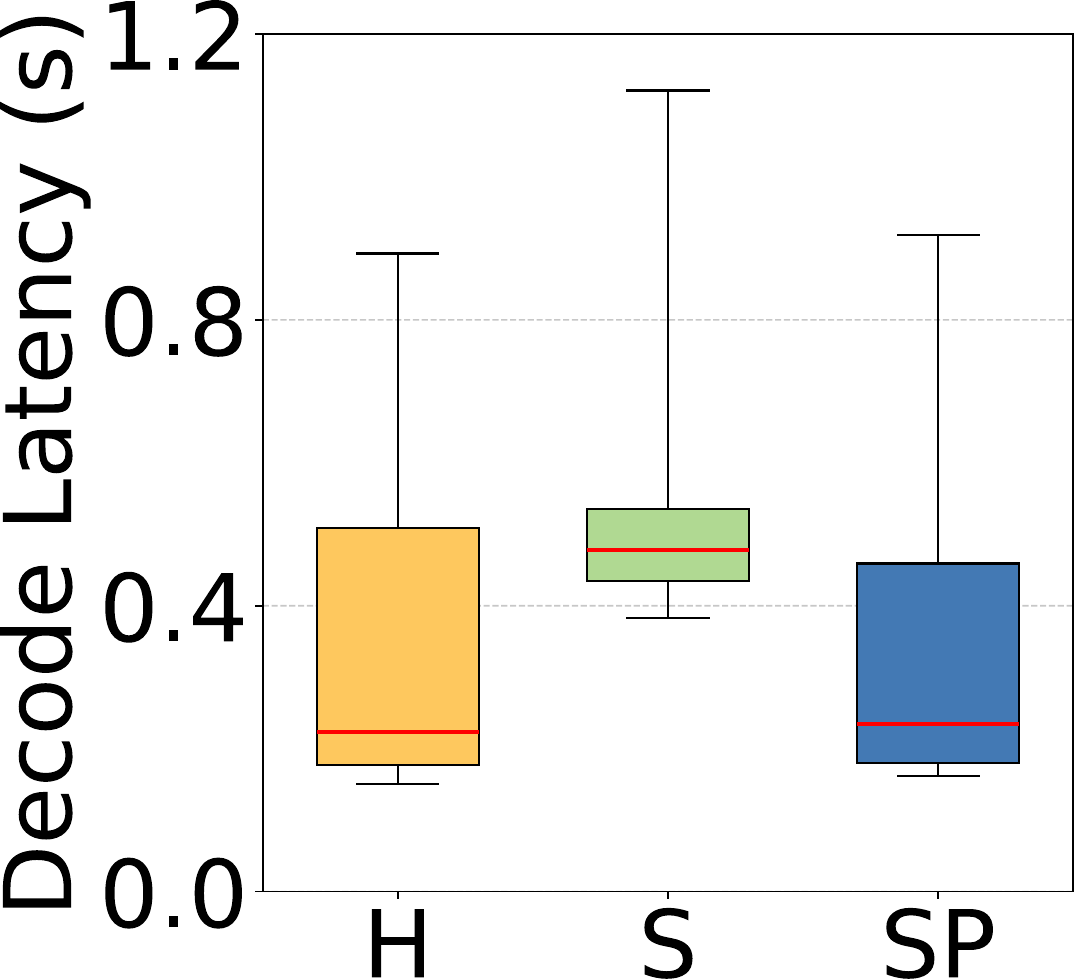}
    \caption{LLaMA-30B}
    \label{fig:sec6-distributed-30b-decode-latency}
    \end{subfigure}
    \begin{subfigure}{0.15\linewidth}
    \centering
    \includegraphics[width=\linewidth]{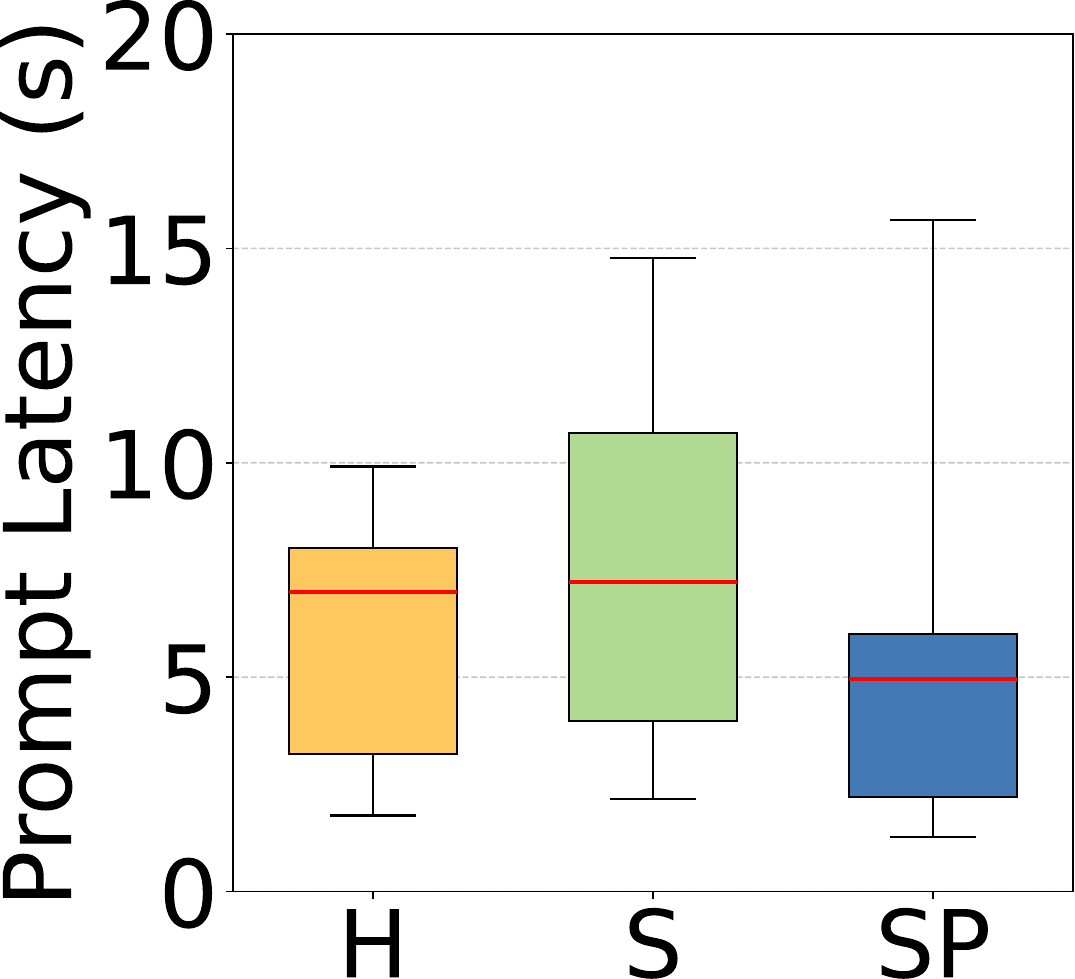}
    \caption{LLaMA-70B}
    \label{fig:sec6-distributed-70b-prompt-latency}
    \end{subfigure}
    \begin{subfigure}{0.14\linewidth}
    \centering
    \includegraphics[width=\linewidth]{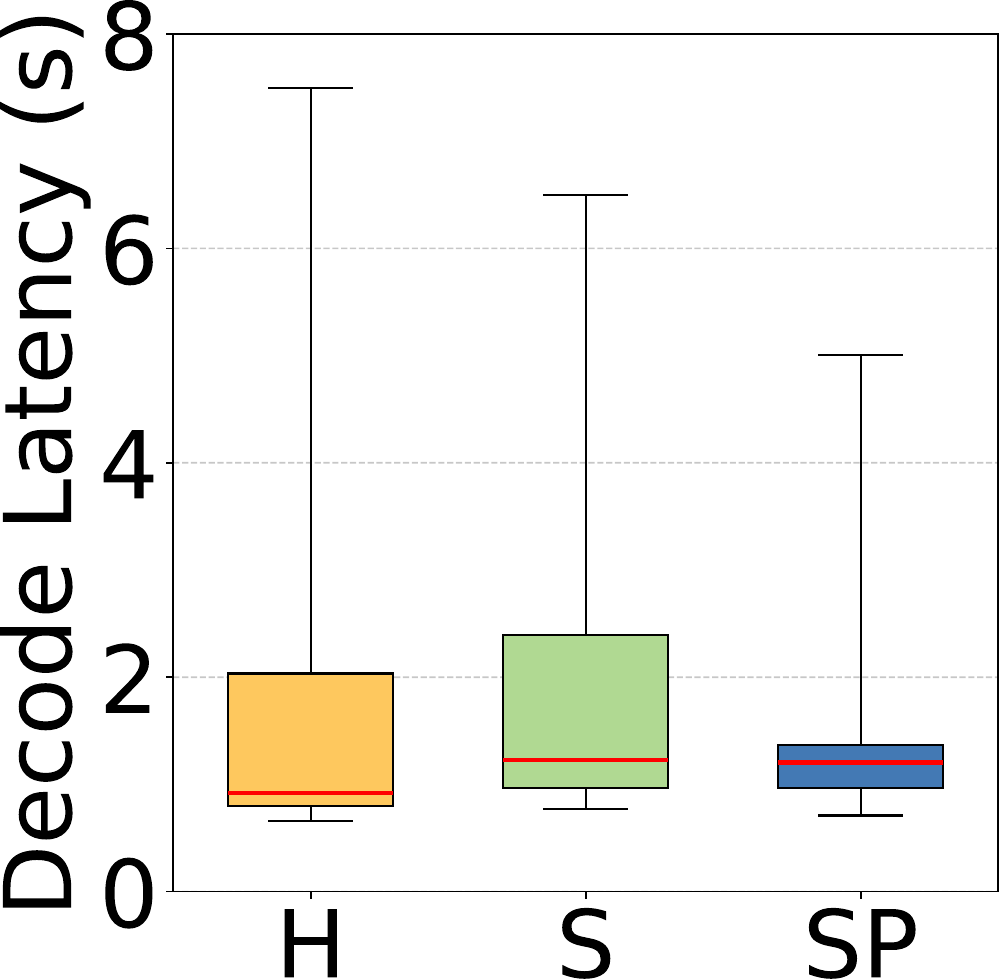}
    \caption{LLaMA-70B}
    \label{fig:sec6-distributed-70b-decode-latency}
    \end{subfigure}
    \vspace{-0.8em}
    \caption{\textbf{Geo-distributed clusters:} \textbf{(a - b)} decode throughput, \textbf{(c - f)} prompt and decode latency for online serving. "H" stands for \sys, "S" stands for Swarm. Box: 25 and 75 percentile. Whisker: 5 and 95 percentile. Red line: median.}
    \vspace{-0.3em}
\label{fig:sec6-geo-distributed-clusters}
\Description{Figure for geo-distributed cluster results.}
\end{figure*}


\subsection{Experiment Setup}
\label{sec6:experiment-setup}
\paragraph{Models.} We evaluate \sys on LLaMA~\cite{touvron2023llama,touvron2023llama2}, a representative and popular open-source Transformer model family. Specifically, we use \camready{LLaMA-1 30B and LLaMA-2 70B} to study the system performance on models of different sizes. We run model inference with half-precision (FP16). \camready{In subsequent sections, we refer to these two models as LLaMA 30B and LLaMA 70B.}

\paragraph{Cluster setup.} {We evaluate with three cluster setups: (1) single cluster (Sec.~\ref{sec6:single-cluster}), (2) geo-distributed clusters (Sec.~\ref{sec6:distributed-clusters}), and (3) high GPU heterogeneity cluster (Sec.~\ref{sec6:cluster_heterogeneity}). First, the \textit{single cluster} setup contains 4 A100 nodes, 8 L4 nodes and 12 T4 nodes connected by 10Gb/s network. \camready{We configure the network bandwidth to 10 Gb/s, as this rate is sufficient to ensure that LLM serving is limited by GPU throughput rather than network capacity.} We allocate the nodes within one region on the Google Cloud. 
Second, the \textit{geo-distributed clusters} contain three clusters with (i) 4 A100 nodes, (ii) 2 L4 nodes + 8 T4 nodes, and (iii) 6 L4 nodes + 4 T4 nodes. Inter-cluster communication has an average bandwidth of 100 Mb/s and an average latency of 50 ms. \camready{We configure the bandwidth to 100 Mb/s to simulate cross-region network limitations, based on our profiling results in Table~\ref{tab:network_condition}.}
Finally, the \textit{high GPU heterogeneity cluster} contains 4 A100 nodes, 6 V100 nodes, 8 L4 nodes, 10 T4 nodes, 4 2$\times$L4 nodes, 6 2$\times$T4 nodes and 4 4$\times$T4 nodes.
The latter two setups are performed in simulation. Our evaluation of the prototype system focuses on single-GPU nodes, as multi-GPU nodes are much more difficult to allocate in the cloud~\cite{thorpe2023bamboo}. Our implementation also works for multi-GPU nodes by leveraging tensor model parallelism across GPUs on the same node as supported by vLLM~\cite{kwon2023vllm}.}
\paragraph{Traces.} The traces we use come from Azure Conversation dataset~\cite{patel2023splitwise}. Fig.~\ref{fig:sec6-azure} shows the length distribution and arrival rate of this dataset. We remove requests with input lengths larger than 2048 or output lengths larger than 1024 to \camready{maintain reasonable runtime memory usage for vLLM}. The pruned dataset contains 16657 requests with an average input length of 763 and an average output length of 232.
{We remark that this dataset is more challenging than the ones used by prior work~\cite{kwon2023vllm} due to longer request length.}
We have two settings for arrival rates. For \textbf{online setting}, we use real arrival rates from Azure Conversation dataset. We scale the average arrival rate to 75\% of the cluster's peak throughput to avoid bursts of requests leading to OOM in the system. 
For \textbf{offline setting}, we allow requests to arrive at the rate needed to fully utilize the cluster. This mimics running offline inference on a dataset. We refer to the two settings as \textit{online} and \textit{offline serving}. 

\paragraph{Experiment duration.} For online setting, we warm up the cluster for 30s and test for 30 minutes. For offline setting, we warm up the cluster for 1 minute and test for 10 minutes. {This amount of time is sufficient for our evaluations and we do not run longer to reduce unnecessary experimental cost.} 

\paragraph{\sys setup.} We allow the MILP solver to search w/ and w/o partial inference and cluster pruning. We also hint the MILP solver with solutions from Petals / Swarm / separate pipelines. Solving times out when the MILP solver does not find better solutions in 10 minutes. The max search budget for each cluster setup is 4 hours on a 14-core CPU. 

\paragraph{Baselines.} To the best of our knowledge, there are no heterogeneous LLM serving systems with model placement and scheduling applicable to our settings. Therefore, we adopt ideas from a heterogeneous LLM training system, Swarm~\cite{ryabinin2023swarm}, to build a competitive heterogeneous baseline. {We also build another baseline that handles GPU heterogeneity by serving one model replica with each type of machine.} We refer to the two baselines as Swarm and separate pipelines (SP). \camready{To ensure a fair comparison, we implemented the SP baseline within our system rather than utilizing other systems.}
{We further compare with the model placement of Petals~\cite{borzunov2022petals} in Sec.~\ref{sec6:model-placement}, which is a decentralized LLM serving system for volunteer computing. We do not compare with Petals in end-to-end serving as it lacks centralized request scheduling.}
{For Swarm, we implemented their model placement and scheduling algorithm in our system, since their original system can not be used for inference. 
We set the number of pipeline stages to the minimum that allows the weakest GPU to hold one stage with half its VRAM. This minimizes the pipeline depth and leaves enough VRAM for KV-cache, both of which are crucial to performance.}


\paragraph{Metrics.} For offline serving, we report average \textit{decode throughput}, which is the number of tokens generated per second. For online serving, we further report average \textit{prompt latency} and \textit{decode latency}, which is the average latency for parsing user input and generating new tokens, respectively. 

\begin{figure}[t]
    \centering
    \begin{subfigure}{0.38\linewidth}
    \centering
    \includegraphics[width=\linewidth]{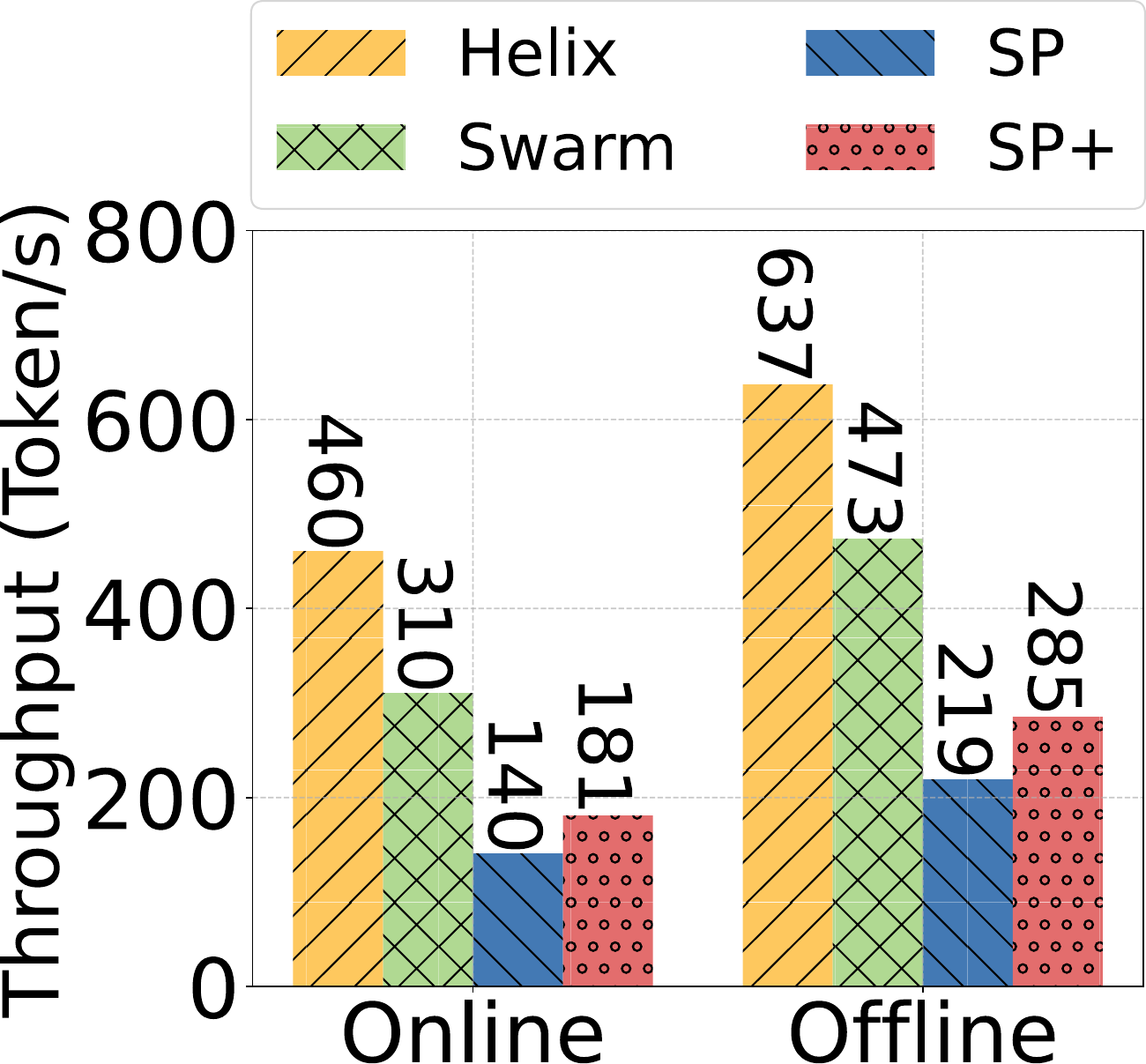}
    \caption{LLaMA-70B}
    \label{fig:sec6-high-hetero-70b-throughput}
    \end{subfigure}
    \begin{subfigure}{0.3\linewidth}
    \centering
    \includegraphics[width=\linewidth]{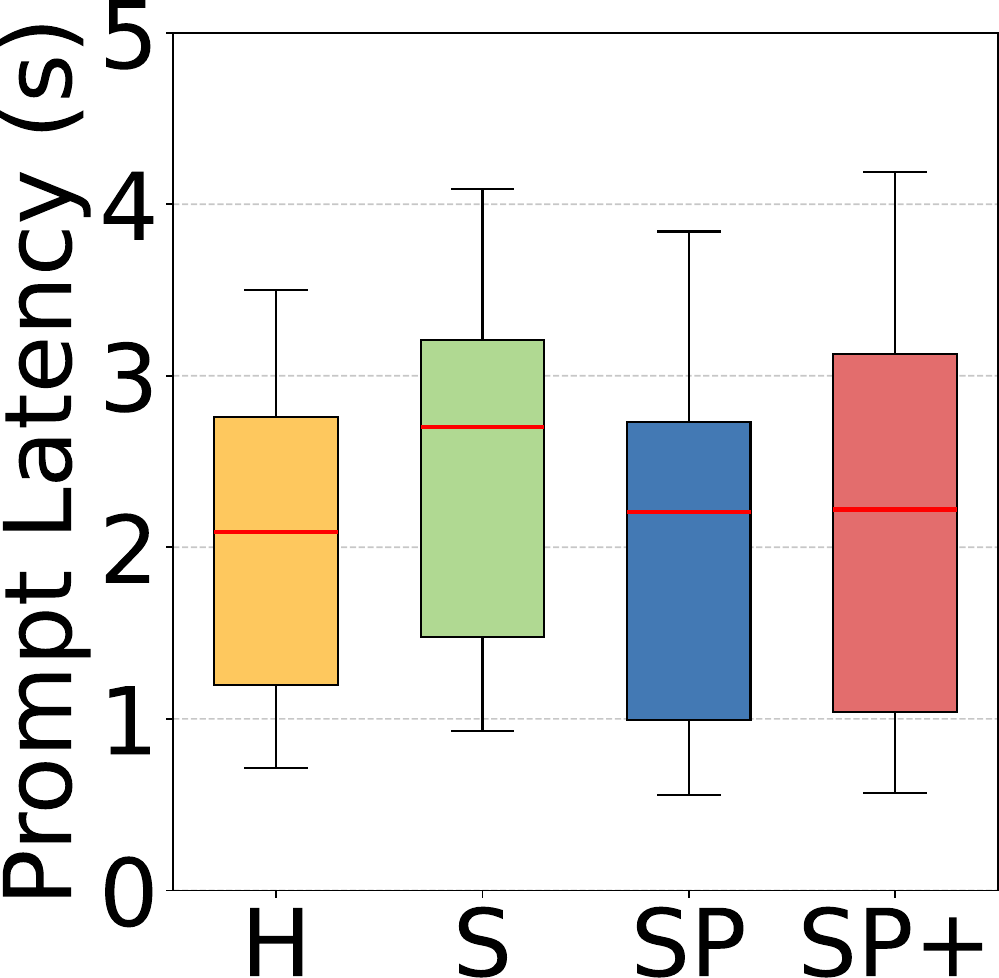}
    \caption{LLaMA-70B}
    \label{fig:sec6-high-hetero-70b-prompt}
    \end{subfigure}
    \begin{subfigure}{0.3\linewidth}
    \centering
    \includegraphics[width=\linewidth]{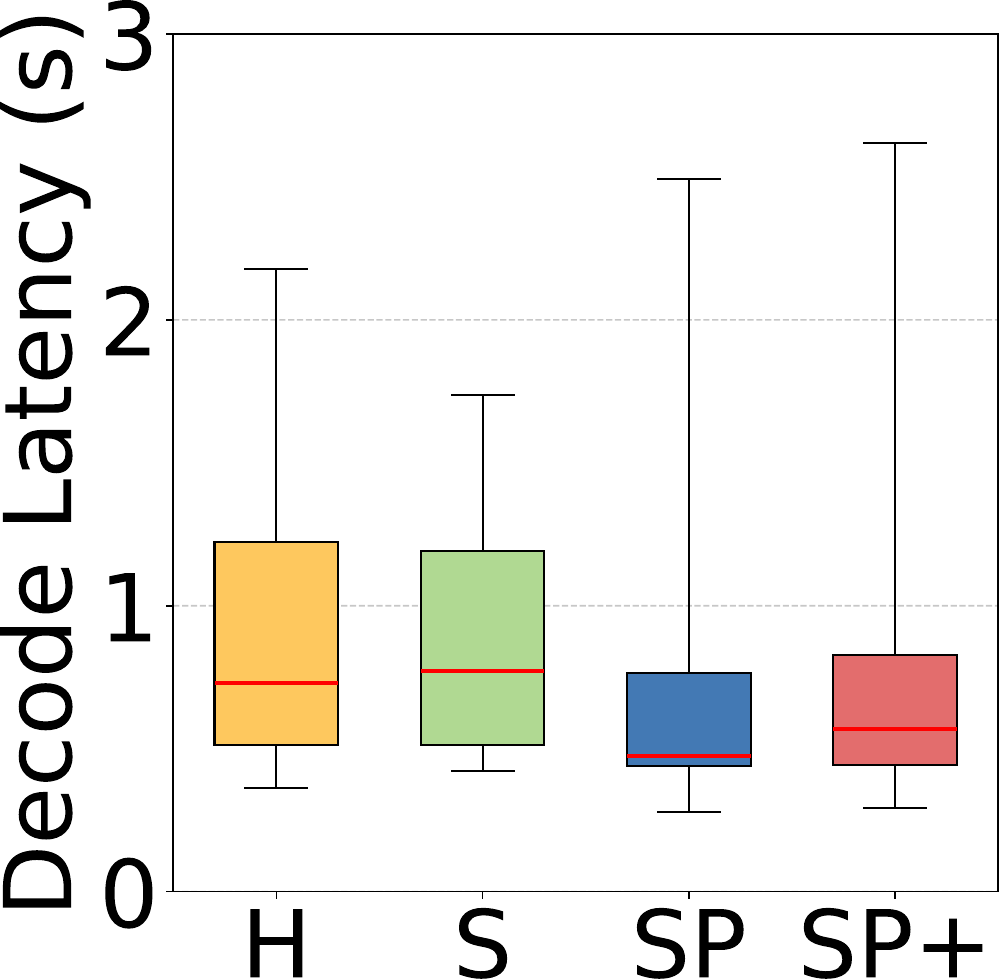}
    \caption{LLaMA-70B}
    \label{fig:sec6-high-hetero-70b-decode}
    \end{subfigure}
    \vspace{-1.5em}
    \caption{\textbf{High GPU-heterogeneity clusters setup:} \textbf{(a)} decode throughput, \textbf{(b - c)} prompt and decode latency for online serving. "H" stands for \sys, "S" stands for Swarm.}
\label{fig:sec6-high-hetero-clusters}
\Description{Figure for high GPU-heterogeneity cluster results.}
\end{figure}

\begin{figure*}[t]
    \centering
    \begin{subfigure}{0.22\linewidth}
    \centering
    \includegraphics[width=\linewidth]{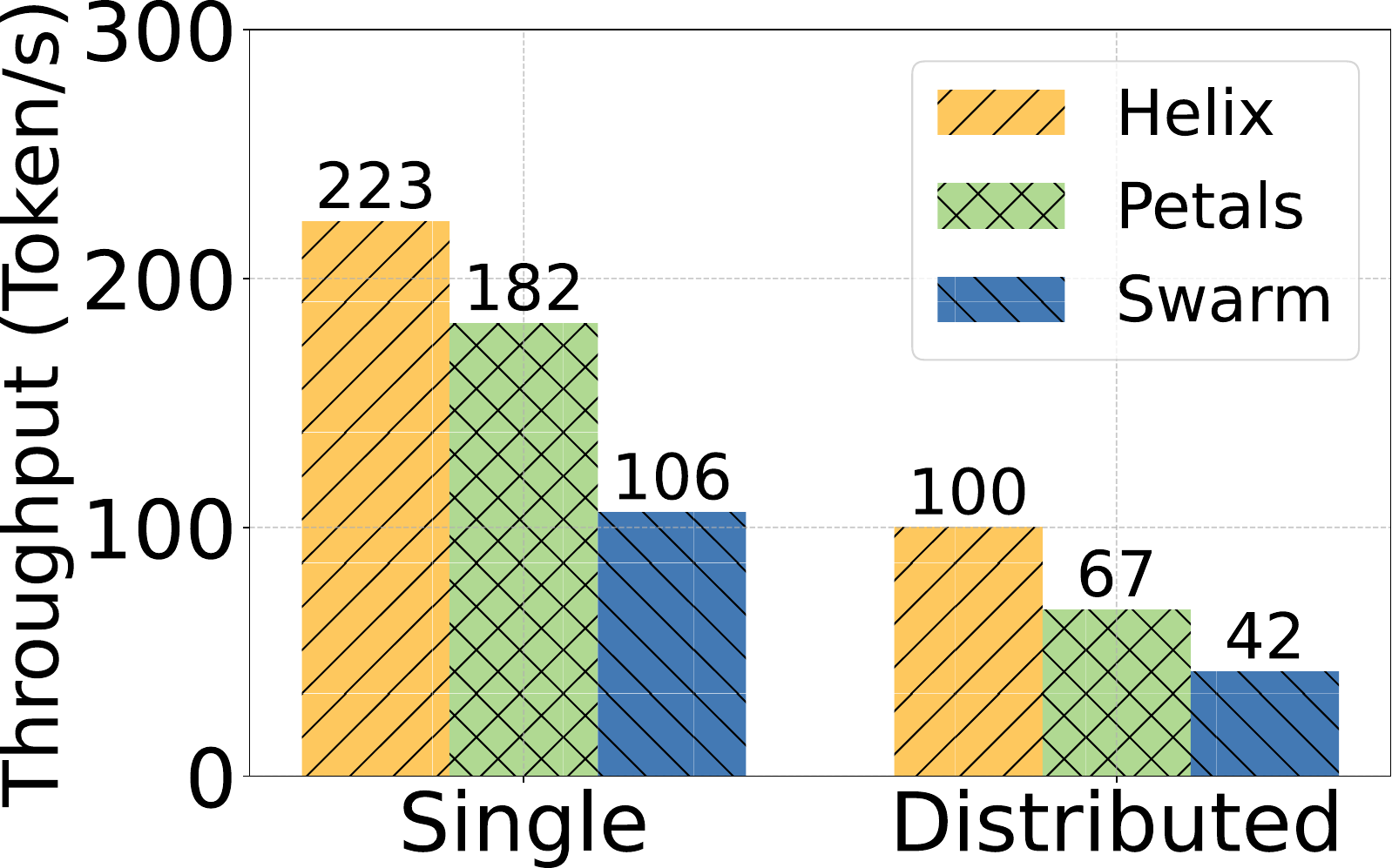}
    \caption{Decode throughput.}
    \label{fig:sec6-model-placement-throughput}
    \end{subfigure}
    \hfill
    \begin{subfigure}{0.76\linewidth}
    \centering
    \includegraphics[width=\linewidth]{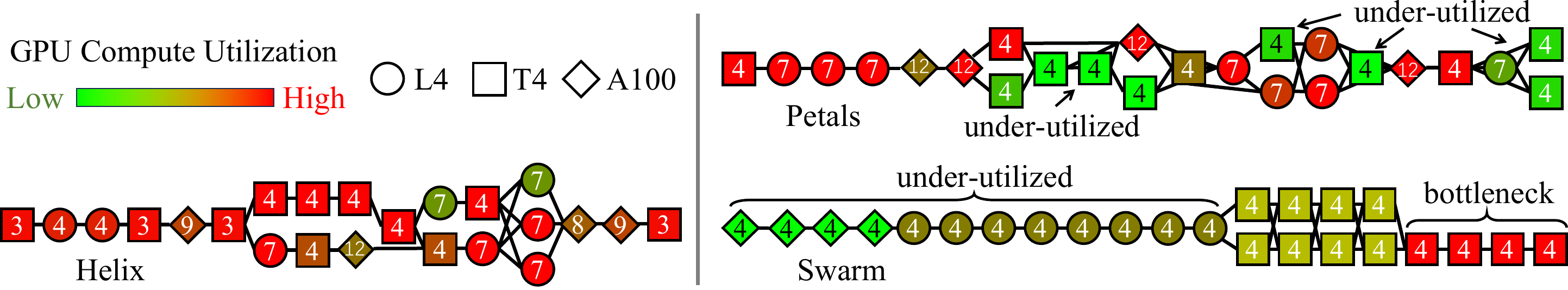}
    \caption{Model placement case study. Numbers on each node represent the number of layers the node holds.}
    \label{fig:sec6-model-placement-case-study}
    \end{subfigure}
    \caption{{\textbf{(a)} Comparing different model placement methods with offline serving of LLaMA 70B. \textbf{(b)} The case study shows model placements for serving LLaMA 70B on the single cluster setup.}}
    \label{fig:sec6-placement}
    \vspace{-0.4em}
    \Description{Figure for model placement.}
\end{figure*}

\subsection{Single Cluster}
\label{sec6:single-cluster}

{This section evaluates \sys with online and offline serving of LLaMA 30B and 70B in the single cluster setup. Fig.~\ref{fig:sec6-single-cluster} shows the results.}
{For LLaMA 30B, each GPU type has enough nodes to serve at least one individual pipeline. 
The best model placement discovered by \sys serves three replicas of the model, each with one type of GPU. As a result, \sys and SP achieve similar performance: \sys has 4\% and 14\% higher decode throughput for offline and online serving because of its better KV-cache utilization; the prompt latency is 2\% lower and decode latency is 10\% higher. Compared with Swarm, \sys achieves 2.14$\times$ and 2.07$\times$ higher decode throughput for offline and online serving. \sys also reduces prompt and decode latency by 32\% and 12\% for online serving. We find that Swarm's model placement introduces a bottleneck and under-utilizes the A100 nodes, which we discuss in more detail in Sec.~\ref{sec6:model-placement}.} \camready{We note that the comparatively higher latency observed in our experiments, relative to other LLM serving systems~\cite{kwon2023vllm, patel2023splitwise}, can be attributed to our use of less powerful GPUs (T4 and L4) in contrast to the A100 and H100 GPUs utilized in other studies.}

{For LLaMA 70B, nodes from a single GPU type can not serve one model replica by themselves while leaving enough VRAM for KV-cache. In this case, SP's throughput significantly decreases. Compared with SP, \sys achieves 1.86$\times$ and 1.69$\times$ higher decode throughput in offline and online serving. The average latency of \sys is higher than SP because SP serves majority of requests with A100 nodes and under-utilizes the L4 and T4 nodes. Compared with Swarm, \sys achieves 1.94$\times$ and 2.00$\times$ higher decode throughput in offline and online serving. \sys's prompt latency is 15\% lower, but decode latency is 16\% higher because of \sys's high GPU utilization. Similar to LLaMA 30B, Swarm under-utilizes A100 nodes because of its model placement.}

{We also conduct these experiments in the simulator (see Sec.~\ref{sec6:implementation-details}) to examine its fidelity. The simulation results are shown alongside the real system evaluation results in Fig.~\ref{fig:sec6-single-cluster}. The average error of decode throughput is lower than 5\%. The average error of prompt and decode latency is lower than 5\% and 4\% respectively. This indicates that our simulator achieves high fidelity and serves the purpose of comparing the performance of different methods.}

\subsection{{Geo-Distributed Clusters}} 
\label{sec6:distributed-clusters}

{This section evaluates \sys for both online and offline serving of LLaMA 30B and 70B in geo-distributed clusters using our high-fidelity simulator in Sec.~\ref{sec6:implementation-details}. Fig.~\ref{fig:sec6-geo-distributed-clusters} shows the results. Although this setup involves the same set of GPUs as the single cluster, the slow inter-cluster network causes all methods to have lower throughput and higher latency.}

{For LLaMA 30B, the best model placement found by \sys still consists of three pipelines served by three types of machines separately. Compared with SP, \sys has 7\% and 10\% higher decode throughput for offline and online serving. The prompt latency is 14\% lower and decode latency is 2\% higher. Compared with Swarm, \sys achieves 2.41$\times$ and 2.33$\times$ higher decode throughput for both online and offline serving. \sys also reduces prompt and decode latency by 66\% and 24\% for online serving.}

{On the other hand, serving LLaMA 70B is more sensitive to slow network because of larger activation size and model depth. \sys reduces network overhead by using model placements with shallower pipeline depth. The model placement found by \sys reduces pipeline depth by 28\% compared to Swarm. It is also 19\% shallower than the one found by \sys when network is fast. \sys's model placement planner balances network overhead with single node's GPU utilization, achieving high throughput and low latency in geo-distributed GPU clusters.
Compared with SP, \sys achieves 1.61$\times$ and 1.79$\times$ higher decode throughput for offline and online serving, respectively. 
SP has lower average latency because it serves most requests with fast GPUs, while 12 T4 nodes are under-utilized because of VRAM constraints. 
Compared with Swarm, \sys achieves 1.92$\times$ and 1.97$\times$ higher decode throughput for offline and online serving. \sys also reduces prompt and decode latency by 21\% and 7\%. We observe severe congestion during offline serving with Swarm. The average prompt latency reaches 71s, which is 7.5$\times$ that of \sys's. We will discuss more about the congestion of Swarm with a case study in Sec.~\ref{sec6:request_scheduling}.}

\subsection{{High GPU-Heterogeneity Cluster}}
\label{sec6:cluster_heterogeneity}

{This section evaluates \sys on a highly heterogeneous GPU cluster with 42 compute nodes and 7 types of GPUs. Fig.~\ref{fig:sec6-high-hetero-clusters} shows the results. In this cluster, V100, T4, and T4$\times$2 nodes cannot form serving pipelines by themselves. 
We report the throughput without those machines for SP. We also try to build a mixed pipeline using those machines and report the number with the mixed pipeline as SP+. Compared with Swarm, SP and SP+, \sys achieves 1.37$\times$, 2.91$\times$, and 2.24$\times$ throughput for offline serving, and 1.48$\times$, 3.29$\times$ and 2.54$\times$ for online serving. \sys also reduces prompt latency by 17\%, 1\% and 7\% respectively. The average decode latency of \sys is slightly higher, because the baselines under-utilize the slow GPUs and serve most requests with fast GPUs with large VRAM. This result indicates that \sys can achieve consistently high performance when there are many types of GPUs in the cluster.}



\begin{figure*}[t]
    \centering
    \hspace*{\fill}
    \begin{subfigure}{0.27\linewidth}
    \centering
    \includegraphics[width=\linewidth]{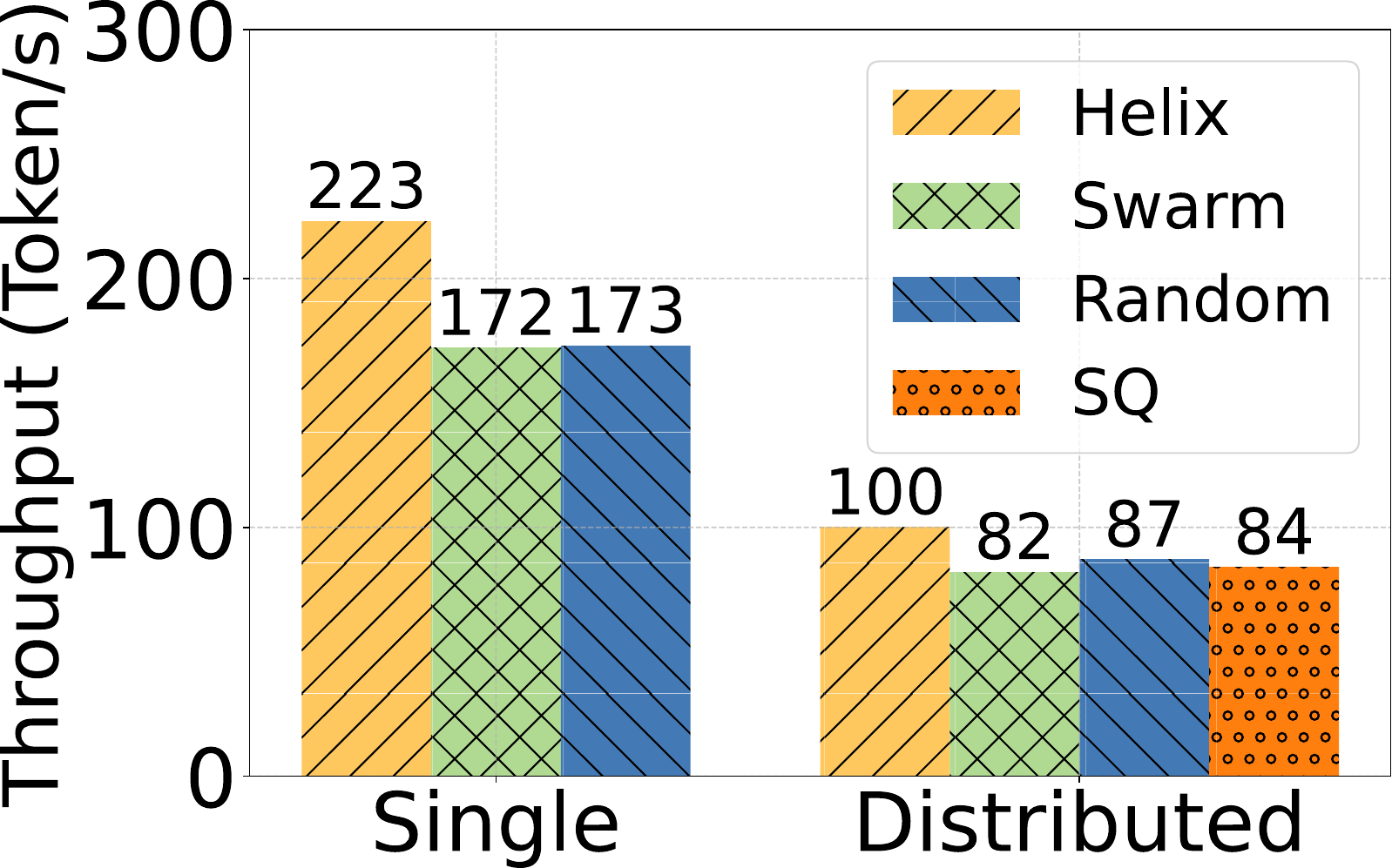}
    \caption{Decode throughput.}
    \label{fig:sec6-scheduling-result}
    \end{subfigure}
    \hfill
    \begin{subfigure}{0.6\linewidth}
    \centering
    \includegraphics[width=\linewidth]{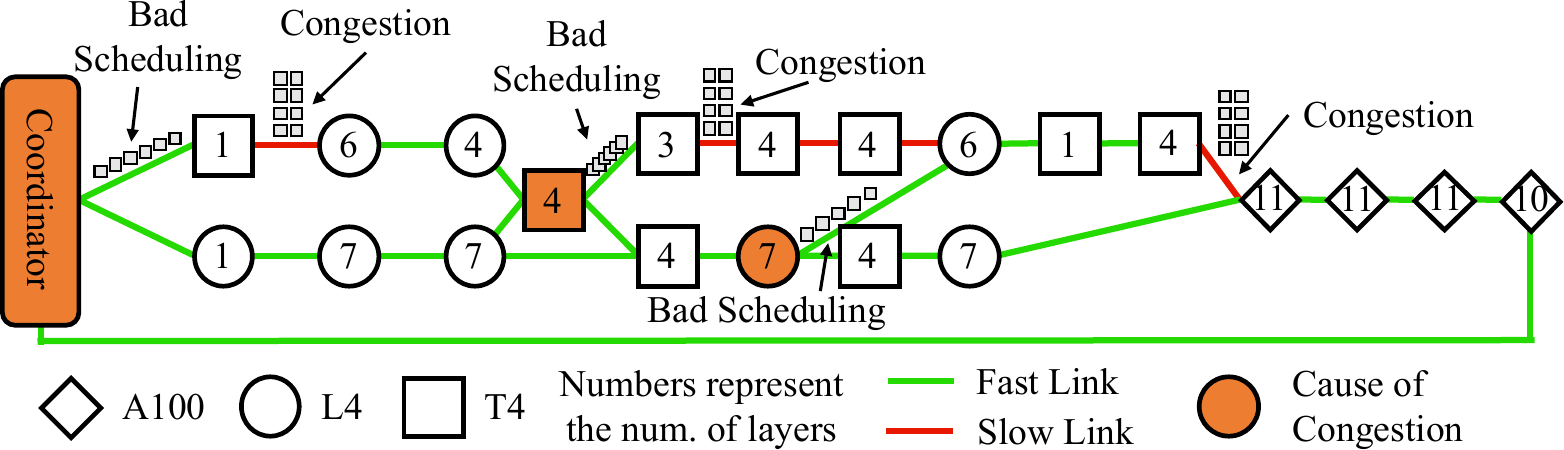}
    \caption{Request scheduling case study.}
    \label{fig:sec6-scheduling-case}
    \end{subfigure}
    \hspace*{\fill}
    \vspace{-0.5em}
    \caption{\camready{\textbf{(a)} Comparing different request scheduling methods with offline serving of LLaMA 70B. \textbf{(b)} A case study that illustrates the congestion in Swarm and random scheduling when serving LLaMA 70B in geo-distributed clusters setup.}
    \vspace{-0.4em}
    }
    \label{fig:sec6-scheduling}
    \Description{Figure for request scheduling.}
\end{figure*}



\subsection{Model Placement Deep Dive}
\label{sec6:model-placement}

In this section, we analyze the impact of different model placement methods on the serving throughput of the cluster. {We evaluate \sys's offline serving performance on both single and geo-distributed clusters, and compare \sys's model placement method with those used in Swarm and Petals. 
Both methods perform model placement using throughput-based heuristics. 
To isolate the effect of model placement, we use \sys's request scheduler for all methods. Fig.~\ref{fig:sec6-model-placement-throughput} shows the decode throughput. Compared with Petals and Swarm, \sys achieves 1.23$\times$ and 2.10$\times$ higher throughput on the single cluster, and 1.49$\times$ and 2.38$\times$ throughput on the geo-distributed clusters. We perform a case study on LLaMA 70B to demonstrate why \sys achieves the best performance.}

\paragraph{Case study: LLaMA 70B - single cluster.}
{Fig.~\ref{fig:sec6-model-placement-case-study} shows the model placement and GPU compute utilization for each method when serving LLaMA 70B on the single cluster. 
Swarm's model placement introduces a bottleneck at the end of its pipeline, where 4 T4 nodes each serves 4 layers. This bottleneck causes GPU under-utilization on A100 and L4 nodes, significantly decreasing the serving throughput. 
For Petals' model placement, 8 T4 nodes and 1 L4 node are under-utilized, which negatively affects the serving throughput. For \sys's model placement, almost all nodes are fully-utilized. The efficient use of GPUs enables \sys to outperform Swarm and Petals by 2.10$\times$ and 1.23$\times$ respectively.}

\subsection{Request Scheduling Deep Dive}
\label{sec6:request_scheduling}
This section analyzes the impact of different request scheduling methods on the serving throughput of the cluster. We evaluate \sys's offline serving performance of LLaMA 70B on both the single cluster and geo-distributed clusters. 
We compare \sys's request scheduler with (1) Swarm, which schedules requests based on real-time throughput of each candidate node, and (2) random scheduling, which randomly chooses a candidate in scheduling. \camready{For the geo-distributed clusters setup, we further compare with Shortest Queue First scheduling (SQ), which always assigns requests to the node with shortest queue.}
To eliminate the impact of model placement, all methods use the model placement found by \sys. {Fig.~\ref{fig:sec6-scheduling-result} shows the decode throughput. Compared with Swarm and random scheduling, \sys achieves 30\% and 29\% higher throughput on the single cluster, and 22\% and 15\% higher throughput on the geo-distributed clusters. \camready{Compared with Shortest Queue First scheduling, \sys achieves 19\% higher decode throughput.} Moreover, runtime monitoring shows that \camready{all three baseline scheduling methods} introduce severe congestion. We illustrate this further in the case study.}

%
%
%
%

\paragraph{Case study: LLaMA 70B - distributed clusters.}
{Fig.~\ref{fig:sec6-scheduling-case} shows the model placement plan found by \sys for serving LLaMA 70B on the geo-distributed clusters.} The plan avoids using slow inter-cluster network connections as much as possible, but a few compute nodes are still connected with low-bandwidth connections. When using Swarm or random scheduling to schedule requests, we observe severe congestion on the three links marked as ``congestion'' in the figure -- prompt phase requests queue up on those links for an average of 5s - 16s before they can be transmitted. We root-cause the nodes responsible for the congestion and mark them orange in the figure. Surprisingly, we find that one congestion is caused by bad scheduling from a node 3 hops away. This verifies the necessity of a global scheduling method that can take both network and compute into account. {We also observe similar congestion when serving LLaMA 70B with Swarm's request scheduling on the model placement it finds for the geo-distributed clusters.}

\begin{table}
    \centering
    \caption{Problem size with and without pruning. var means variables, and cstr means constraints. 
    \vspace{-0.4em}
    }
    \label{tab:sec6-prune}
    \small
    \begin{tabular}{|c|c|c|}
    \hline
    {\bf Problem size} & {\bf With pruning} & {\bf Without pruning}\\
    \hline
    24-node & 876 var 1122 cstr & 1376 var 1848 cstr \\
    \hline
    42-node & 2144 var 2772 cstr & 4004 var 5502 cstr \\
    \hline
    \end{tabular}
\end{table}

\begin{figure}[t]
    \begin{subfigure}{0.48\linewidth}
    \centering
    \includegraphics[width=\linewidth]{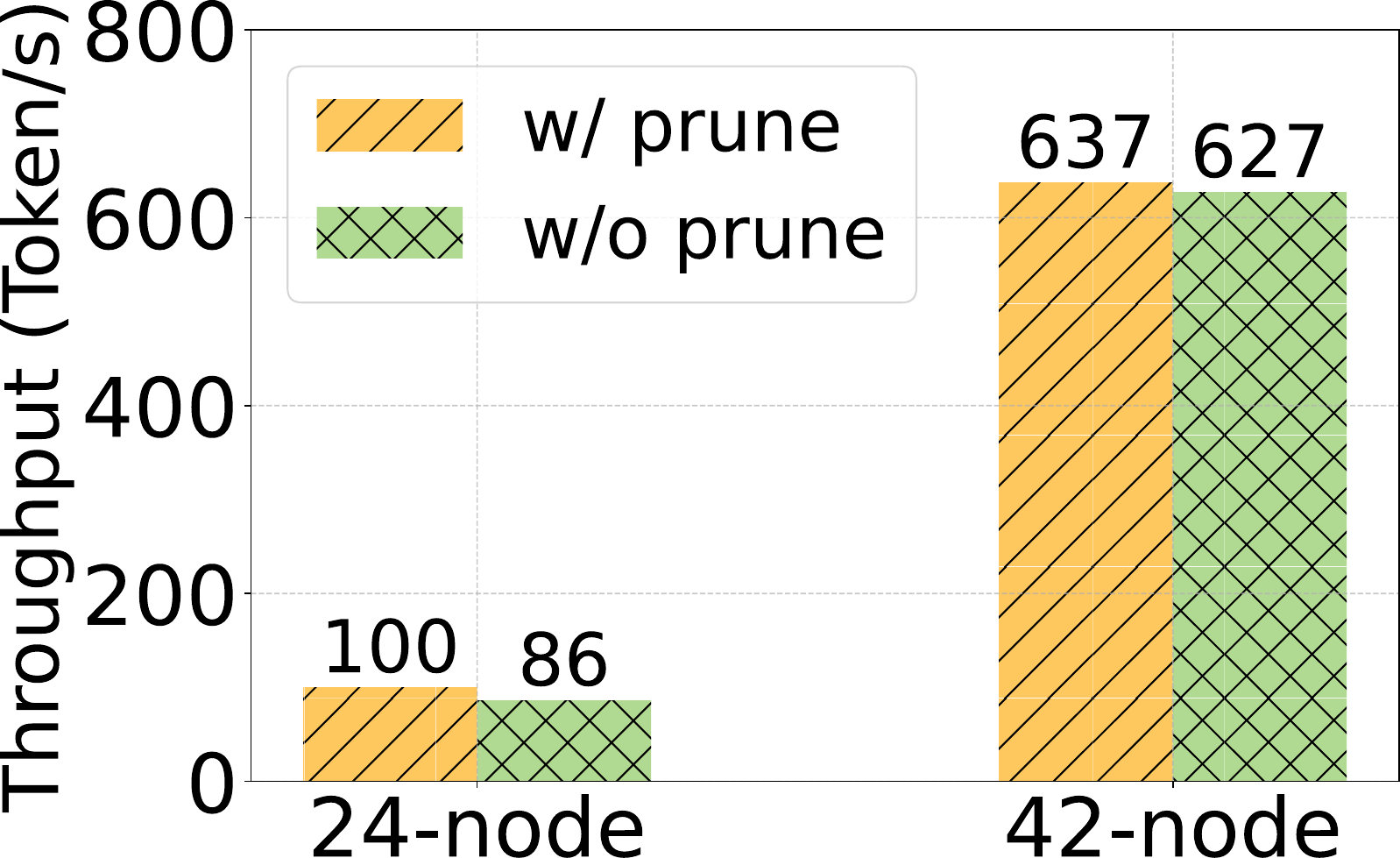}
    \caption{Cluster pruning.}
    \label{fig:sec6-ablation-prune}
    \end{subfigure}
    \begin{subfigure}{0.48\linewidth}
    \centering
    \includegraphics[width=\linewidth]{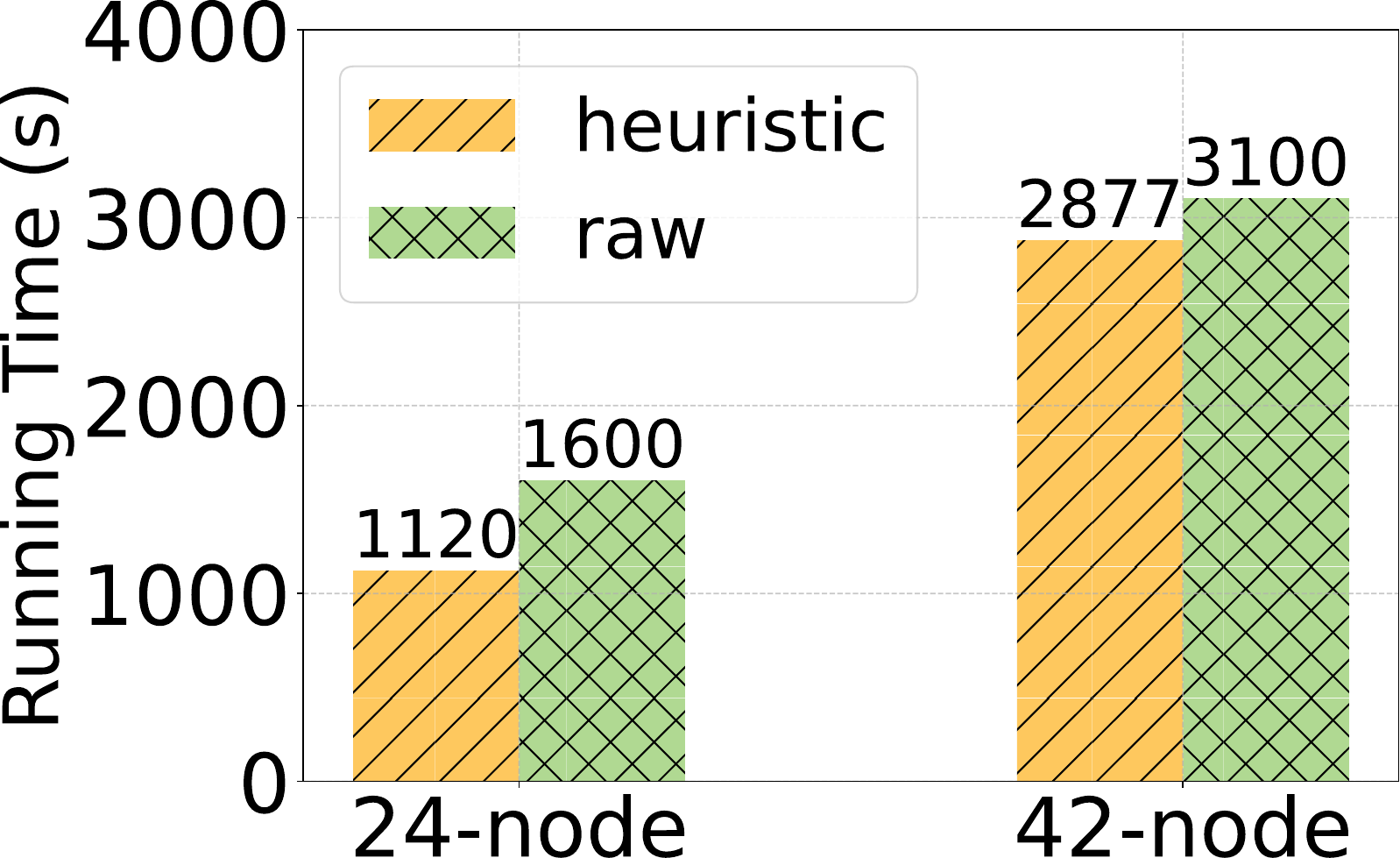}
    \caption{Initial values.}
    \label{fig:sec6-ablation-heuristics}
    \end{subfigure}
    \vspace{-0.4em}
    \caption{{Ablation study on MILP optimization.}
    \vspace{-0.8em}
    }
    \label{fig:sec6-sensitivity}
    \Description{Figure for ablation study.}
\end{figure}

\subsection{Ablation Study on Optimization}
\label{sec6:ablation}
This section performs an ablation study on the two MILP optimizations introduced in Sec.~\ref{sec4:milp-analysis}. {We evaluate offline serving of LLaMA 70B on the geo-distributed and highly heterogeneous clusters, referred to as the 24-node and 42-node settings respectively.}

\paragraph{Cluster pruning.} When cluster pruning is enabled, we prune network connections such that the average degree of each node is 12, which is sufficient for LLM inference systems as we discuss below. Enabling cluster pruning removes 50\% and 72\% network connections for 24 and 42-node settings. Table.~\ref{tab:sec6-prune} shows that pruning reduces problem size by 36\% and 46\% for the two settings. Fig.~\ref{fig:sec6-ablation-prune} shows that \sys achieves 16\% and 2\% higher decode throughput when using the model placement found with cluster pruning. We note that the amount of speed-up achieved would vary depending on the specific instance of the MILP problem at hand. Pruning slow network connections does not harm throughput because network connections used in serving is very sparse -- usually each node only communicates with a few other nodes. Also, there are many equivalent model placements that can achieve the same throughput. Pruning the cluster very likely keeps some of these placements still valid. It makes the search for these placements easier with limited optimization time, as the problem size (and solution space size) is reduced.


\paragraph{Initial values.} We compare the performance of running \sys's model placement planner starting from solutions of heuristic methods and from default values. Since the best model placements found are the same, we compare the wall clock time to find the placement. Fig.~\ref{fig:sec6-ablation-heuristics} shows that running MILP from heuristic solutions takes 43\% and 8\% less time for the 24- and 42-node setup. We note that the speed-up achieved would vary depending on the specific instance of the MILP problem at hand. The results show that starting from heuristic solutions accelerates model placement in \sys.

\subsection{Model Placement Quality}

\begin{figure}
    \centering
    \includegraphics[width=0.6\linewidth]{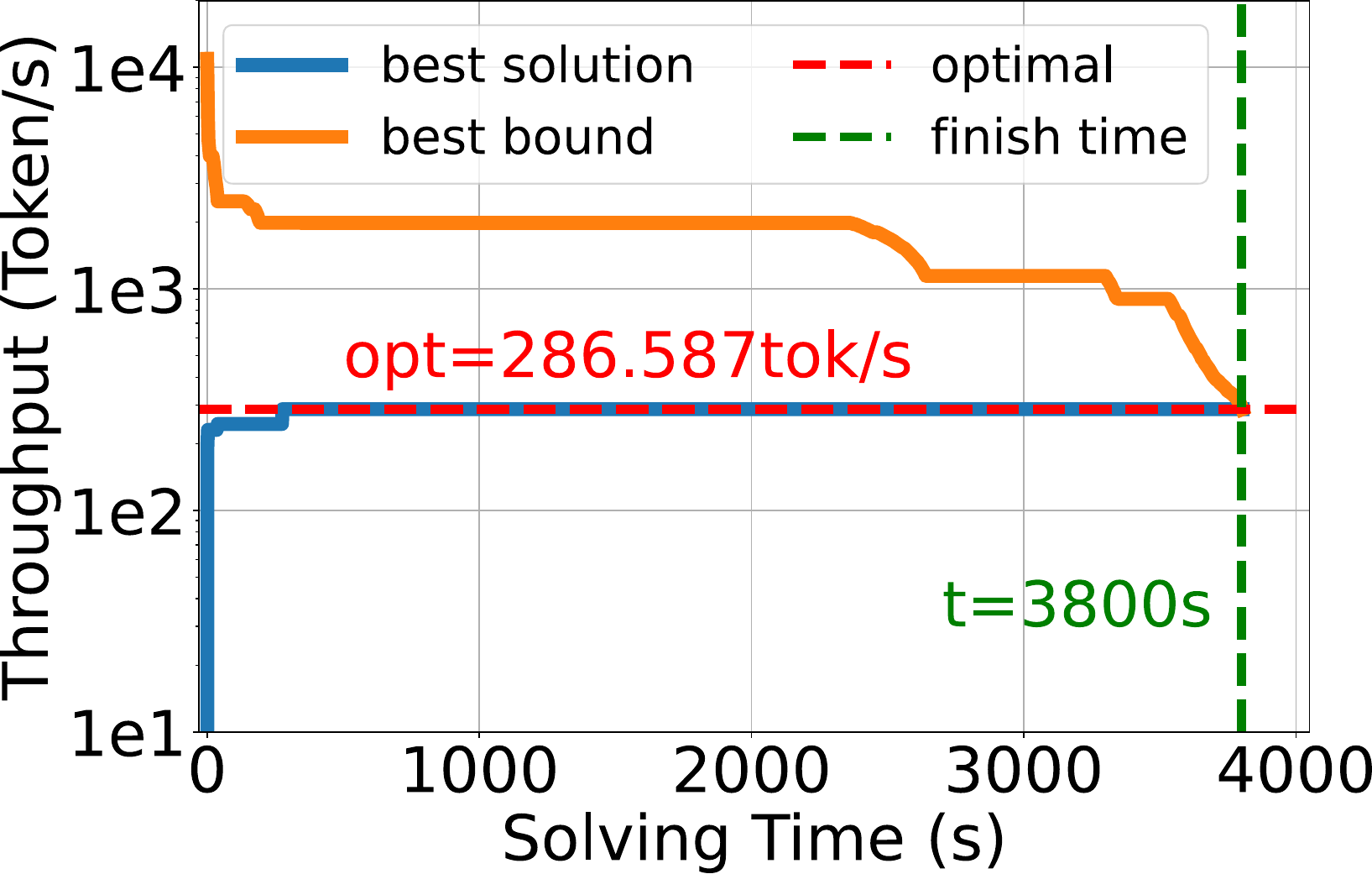}
    \vspace{-0.4em}
    \caption{\camready{Best solution and best upper bound found by the MILP solver relative to solving time. The red dotted line marks the optimal throughput for this cluster.}}
    \vspace{-0.4em}
    \label{fig:sec6-solution-optimality}
    \Description{Figure for model placement quality.}
\end{figure}

\camready{This section evaluates the quality of model placements found by \sys relative to the MILP solving time. In this evaluation, we run \sys to find the optimal model placement for serving LLaMA 30B in a cluster with 4 L4 and 6 T4 machines. We record the best model placement as well as the best upper bound found by the MILP solver that we used (Gurobi) during the solving process. The best upper bound represents the best possible objective value that could be achieved for the MILP problem and will gradually become tighter as the solver explores more nodes and adds cutting planes.
Fig.~\ref{fig:sec6-solution-optimality} shows the quality of the best model placement and the best upper bound relative to the solving time. Results show that \sys finds the optimal solution in less than 5 minutes, but it takes the solver more than one hour to enumerate all possible solutions and confirm the optimality. This indicates that we can early-stop the solving process, as high-quality solutions emerge in the early stages of computation.}

\section{Related Work}
\label{sec:related_work}



\paragraph{Machine Learning Model Serving} There are a large number of works for serving machine learning models, discussing aspects including system implementation~\cite{olston2017tensorflow,nvidiaTriton}, model placement~\cite{li2023alpaserve,pope2023efficiently,shen2019nexus,wu2023transparent}, request scheduling~\cite{gujarati2020serving,zhang2019mark,shen2019nexus}, and tail-latency mitigation~\cite{kosaian2019parity,kosaian2020learning,narra2020collage}. 
However, due to LLM's unique auto-regressive execution paradigm, these approaches fail to efficiently serve LLMs. 
Instead, many recent LLM-specific systems tackle the unpredictable execution time and high memory consumption in LLM serving. 
Orca~\cite{yu2022orca} proposed iteration level scheduling to release resources once a request is finished. 
vLLM~\cite{kwon2023vllm} introduced PageAttention to further reduce the memory consumption of each request by allocating exact number of pages it requires. 
Speculative Inference~\cite{leviathan2023fast,miao2023specinfer} applies a small model to predict multiple output tokens, and verify them in a single iteration. 
Splitwise~\cite{patel2023splitwise} and DistServe~\cite{zhong2024distserve} found that disaggregating the prompt and decode phase can improve the throughput, since the two phases have different workload characteristics. 
Sarathi~\cite{agrawal2023sarathi} introduced chunked prefill, which allocates a budget to the prompt phase to make each microbatch's workload balanced, minimizing pipeline bubble. 
All above works are orthogonal to our work and can be integrated into our system, since our focus is on the cluster heterogeneity.

\paragraph{ML workloads on heterogeneous clusters.} 
Several methods have utilized heterogeneous GPUs for ML tasks. 
Some of them~\cite{park2020hetpipe, jia2022whale} co-design the model partition and placement on a heterogeneous cluster but assume a uniform network bandwidth. 
Learning\@home~\cite{ryabinin2020towards} and DeDLOC~\cite{diskin2021distributed} studied the network-aware routing on a decentralized cluster but only considers either data or pipeline parallelism individually. 
SWARM~\cite{ryabinin2023swarm}, as discussed in Sec.~\ref{sec2:transformer-and-llm-serving}, optimized the pipeline communication in a heterogeneous network. However, it schedules only by the next stage's metadata, lacking a global view. 
There are also several efforts on using approximations to reduce network communication~\cite{wang2023cocktailsgd} or synchronization~\cite{hsieh2017gaia}.
Most of them focus on model training. In model inference, especially LLMs, serving with heterogeneous and geo-distributed GPUs is not well studied. 
SkyPilot~\cite{yang2023skypilot} and Mélange~\cite{griggs2024m} select the best type of GPUs for a request, but each request is served by a single GPU type.
Petals~\cite{borzunov2022petals}, as discussed in Sec.~\ref{sec2:transformer-and-llm-serving}, studies a decentralized pipeline parallel setup. It designs a greedy model allocation and request scheduling for a dynamical device group, losing optimizing opportunities for a fixed device group.
HexGen~\cite{jiang2023hexgen} is a concurrent work on LLM serving in heterogeneous clusters. It is based on fixed pipelines (see Sec.~\ref{sec5:scheduler-design}) and adopts heuristic-based methods to search for the optimal model placement. In comparison, our Max Flow formulation and per-request pipeline is more flexible than HexGen and our MILP formulation can guarantee the optimal solution.

\paragraph{Scheduling for Heterogeneous Resources} \camready{There exists extensive research on scheduling algorithms for heterogeneous resources. For example, energy-aware scheduling~\cite{linux_energy_sched} in the Linux kernel schedules tasks for heterogeneous CPU topologies. There are also several works~\cite{topcuoglu2002performance, radulescu2000fast, bajaj2004improving} that focus on general scheduling in heterogeneous clusters. Due to the auto-regressive nature of LLM inference, these methods cannot be directly used for heterogeneous LLM serving.}

\section{Conclusion}
\label{sec:conclusion}

This paper presents \sys, the first high-throughput, low-latency LLM serving engine for heterogeneous GPU clusters, with guaranteed optimal solution that maximizes throughput. \sys formulates and solves the model placement and request scheduling as a Max-Flow problem. 
Compared to existing solutions, \sys achieves significant improvements in throughput and latency. 


\section*{Acknowledgment}
\camready{We thank the anonymous reviewers and our shepherd Íñigo Goiri for their valuable feedback and constructive suggestions that helped improve this paper. We also express our gratitude to the Google Cloud Innovator program for providing the machines on Google Compute Engine for our experiments.
This work was supported in part by a Sloan Foundation Fellowship and a VMware Systems Research Award. This work was also partially supported by the National Science Foundation under grant numbers CNS-2147909, CNS-2211882, and CNS-2239351, along with gift awards from Amazon and Meta.}

\appendix
\section{Artifact}

\subsection{Abstract}

Our artifacts include Helix's simulator and prototype system for distributed LLM serving in heterogeneous and geo-distributed clusters. The code implements key algorithms including the MaxFlow-based LLM serving formulation, MILP-based model placement planner, and per-request pipeline scheduler. We also provide comprehensive documentation and scripts for environment setup and system execution from scratch. The artifacts contain the identical simulator and prototype system used in Helix's evaluation in the paper. The simulator can run on a single machine, while the prototype system requires a cluster deployment.


\subsection{Artifact check-list (meta-information)}


{\small
\begin{itemize}
  \item {\bf Algorithm: } MaxFlow-based LLM serving formulation; MILP-based model placement planner; per-request pipeline scheduler
  \item {\bf Compilation: } The simulator is implemented in Python. The prototype system's inter-node communication framework is written in C++ (recommended to compile with GCC 13.2) and includes an automated compilation script with CMake, while its remaining components are implemented in Python.
  \item {\bf Model: } We use LLaMa-2 70B as the test workload. The prototype system operates with dummy weights, requiring only the model architecture specification (which is provided in our code repository).
  \item {\bf Data set: } We use the Azure Conversation Dataset as our evaluation traces, with a pre-parsed version included in our code repository.
  \item {\bf Run-time environment: } For the simulator, we recommend using Python 3.10, while it can also support other recent Python versions. The simulator is not sensitive to OS versions. For the prototype system, we recommend using Ubuntu 24.04 LTS to setup the environment. We also recommend using conda to isolate the run-time environment for both systems. For detailed software dependencies, please refer to Sec.~\ref{software}.
  \item {\bf Hardware: } The simulator runs on a single machine without specific hardware requirements. The prototype system runs on a cluster of 24 machines with NVIDIA GPUs. Please refer to Sec.~\ref{hardware} for detailed hardware requirements.
  \item {\bf Metrics: } We evaluate using the same metrics as presented in the paper: decoding throughput, decoding latency, and prompt processing latency.
  \item {\bf Output: } Results are logged to both terminal output and files. We include expected results in the code repository.
  \item {\bf How much disk space required (approximately)?: } The total size of all log files is around 300 MB.
  \item {\bf How much time is needed to prepare workflow (approximately)?: } The environment setup takes around 1 - 2 hours.
  \item {\bf How much time is needed to complete experiments (approximately)?: } Functionality evaluation takes around 2 hours. Reproducibility evaluation takes around 16 hours. (The part that needs the whole 24 machine cluster is around 4 hours. \textbf{Namely, the sections that need the whole cluster are part of Sec 6.3, 6.6 and 6.7.})
  \item {\bf Publicly available?: } Yes. DOI: \href{https://doi.org/10.5281/zenodo.14037926}{10.5281/zenodo.14037926}.
  \item {\bf Code licenses (if publicly available)?: } Apache 2.0
\end{itemize}
}

\subsection{Description}

\subsubsection{How to access}
The artifact is publicly available at \href{https://github.com/Thesys-lab/Helix-ASPLOS25}{https://github.com/Thesys-lab/Helix-ASPLOS25}. It is also archived as \href{https://doi.org/10.5281/zenodo.14037926}{10.5281/zenodo.14037926}. The file size is around 300 MB. Please refer to our Github repository for the latest version.


\subsubsection{Hardware dependencies}
\label{hardware}

The simulator runs on a single machine without specific hardware requirements. We recommend at least 32 GB of memory for simulating large clusters to prevent out-of-memory issues. The prototype system requires cluster deployment - our example configuration uses 24 machines, consisting of 4 machines with 1×A100-40GB, 8 machines with 1×L4, and 12 machines with 1×T4. We recommend to set up each machine with at least 16 CPU cores and 128 GB memory to avoid out-of-memory issues. The network connection between machines should ideally be at least 10 Gbps, with latency of approximately a few milliseconds. Usually machines in the same region from common cloud providers can meet the network requirements. For functionality test purposes, it is also possible to use machines from different regions.

\subsubsection{Software dependencies}
\label{software}

We recommend running the simulator with Python 3.10. It relies on networkx, matplotlib and gurobipy. The MILP-based model placement planner uses Gurobi as the MILP solver. Running the example in the code base does not require additional licenses. However, if you want to run model placement for larger clusters, it is necessary to acquire a Gurobi license that does not limit problem size. We recommend running the prototype system on Ubuntu 24.04 LTS and with Python 3.10. To build the inter-node communication framework, you need to install build-essential, cmake, libzmq, cppzmp and pybind11. To run the prototype system, you need to install CUDA 12.6 and vLLM 0.4.0.post1. For a step-by-step guide to setting up the environment and running the experiments, please refer to \href{https://github.com/Thesys-lab/Helix-ASPLOS25/blob/master/readme.md}{https://github.com/Thesys-lab/Helix-ASPLOS25/blob/master/readme.md}.

\subsubsection{Data sets}
We use the Azure Conversation Dataset as our evaluation traces, with a pre-parsed version included in our code repository.

\subsubsection{Models}
We use LLaMa-2 70B as the test workload. The prototype system operates with dummy weights, requiring only the model architecture specification.

\subsection{Detailed Steps for Reproducing Results}
We provide an example to show the functionality of our system, please refer to \href{https://github.com/Thesys-lab/Helix-ASPLOS25/blob/master/readme.md}{https://github.com/Thesys-lab/Helix-ASPLOS25/blob/master/readme.md}. We also provide another example to reproduce all result we got in the paper, please refer to \href{https://github.com/Thesys-lab/Helix-ASPLOS25/blob/master/artifact_evaluation/ae_readme.md}{https://github.com/Thesys-lab/Helix-ASPLOS25/blob/ master/artifact$\_$evaluation/ae$\_$readme.md}

\bibliographystyle{plain}
\bibliography{references}

\end{document}